\documentclass[aps,prb,floatfix,twocolumn,superscriptaddress,nofootinbib]{revtex4}

\usepackage{amsmath,color,graphicx}

\newcommand{\TC}{T$_{\rm C}$}

\newcommand{\TF}{T$_{\rm F}$}
\newcommand{\TK}{T$_{\rm K}$}

\begin{document}

\title{Spin Textures induced by Quenched Disorder in a Reentrant Spin Glass: Vortices versus "Frustrated" Skyrmions}
\author{I. Mirebeau}\email[]{isabelle.mirebeau@cea.fr}\affiliation{Laboratoire L\'eon Brillouin, CEA, CNRS, Universit\'e Paris-Saclay, CEA Saclay 91191 Gif-sur-Yvette, France}
\author{N. Martin}\affiliation{Laboratoire L\'eon Brillouin, CEA, CNRS, Universit\'e Paris-Saclay, CEA Saclay 91191 Gif-sur-Yvette, France}
\author{M. Deutsch}\affiliation{Universit\'e de Lorraine, CNRS, CRM2, Nancy, France}
\author{L. J. Bannenberg}\affiliation{Faculty of Applied Science, Delft University of Technology, 2629 JB Delft, the Netherlands}
\author{C. Pappas}\affiliation{Faculty of Applied Science, Delft University of Technology, 2629 JB Delft, the Netherlands}
\author{G. Chaboussant}\affiliation{Laboratoire L\'eon Brillouin, CEA, CNRS, Universit\'e Paris-Saclay, CEA Saclay 91191 Gif-sur-Yvette, France}
\author{R. Cubitt}\affiliation{Institut Laue Langevin, BP156, F-38042 Grenoble France}
\author{C. Decorse}\affiliation{ICMMO, Universit\'e Paris-Sud, Universit\'e Paris-Saclay, F-91405 Orsay, France}
\author{A. O. Leonov}\affiliation{Chiral Research Center, Hiroshima University, Higashi-Hiroshima, 739-8526, Japan}\affiliation{Department of Chemistry, Faculty of Science, Hiroshima University Kagamiyama, Higashi Hiroshima, Hiroshima 739-8526, Japan}

\date{\today}

\begin{abstract}
Reentrant spin glasses are frustrated disordered ferromagnets developing vortex-like textures under an applied magnetic field. Our study of a Ni$_{0.81}$Mn$_{0.19}$ single crystal by small angle neutron scattering clarifies their internal structure and shows that these textures are randomly distributed. Spin components transverse to the magnetic field rotate over length scales of 3-15 nm in the explored field range, decreasing as field increases according to a scaling law. Monte-Carlo simulations reveal that the internal structure of the vortices is strongly distorted and differs from that assumed for "frustrated" skyrmions, built upon a competition between symmetric exchange interactions. Isolated vortices have small non-integer topological charge. The vortices keep an anisotropic shape on a 3 dimensional lattice, recalling "croutons" in a "ferromagnetic soup". Their size and number can be tuned \emph{independently} by the magnetic field and concentration $x$ (or heat treatment), respectively. This opens an original route to understand and control the influence of quenched disorder in systems hosting non trivial spin textures.

\end{abstract}


\maketitle

Disorder plays a central role in the advent of the most spectacular quantum phenomena observed in condensed matter. The quantum Hall effect observed in a two-dimensional (2d) electron gas\cite{Klitzing1980,Thouless1982}, the two-current character of the resistivity in impurity-containing ferromagnetic metals\cite{Fert1968} leading to giant magneto-resistance \cite{Baibich1988} or the dissipationless conduction observed in the mixed state of type II superconductors\cite{LeDoussal1998,Klein2001} are prominent examples. Frustrated ferromagnets represent another type of playground to study the influence of disorder. Such systems show competing ferromagnetic (FM)/antiferromagnetic (AFM) interactions combined with atomic disorder. The influence of quenched disorder, when treated in a mean field model with infinite range interactions\cite{Sherrington1975,Gabay1981}, leads to a canonical spin glass (SG) when the average interaction $\bar{J}$ is smaller than the width of the interaction distribution or to a reentrant spin glass (RSG) otherwise. Here, we focus on the FM case ($\bar{J} > 0$) of the RSGs where vortex-like textures are stabilized under an applied magnetic field at low temperature. We study their morphology and spatial organization by combining neutron scattering experiments on a Ni$_{0.81}$Mn$_{0.19}$ single crystal and Monte Carlo simulations. We compare them with those expected for skyrmions built upon a competition between symmetric exchange interactions. Altogether, our study shows that one can \emph{independently} tune the number and size of vortex textures in frustrated disordered magnets with the magnetic field, heat treatment and concentration of magnetic species. It provides clues to control and use the influence of quenched disorder in frustrated ferromagnets and skyrmion-hosting systems in bulk state. 

\section{Reentrant spin glasses and "frustrated" skyrmions}

As a common feature, RSGs show three successive phase transitions upon cooling: a paramagnetic to FM transition at \TC\ followed by transitions towards two mixed phases at \TK\ and \TF . Below the canting temperature \TK\ spin components $\mathbf{m}_{\rm T}$ transverse to the longitudinal magnetization $\mathbf{m}_{\rm L}$ start to freeze. The lower temperature \TF\ marks the onset of strong irreversibilities of $\mathbf{m}_{\rm L}$.  In this picture, the ferromagnetic long range order of $\mathbf{m}_{\rm L}$ is preserved in the RSG down to T $\rightarrow 0$\,K. The phase diagram (T, $x$) where $x$ is a parameter tuning the distribution of interactions shows a critical line between SG and RSGs ended by a multicritical point at $x_{\rm C}$ where all phases  collapse\cite{Gabay1981}. Metallic ferromagnetic alloys with competing nearest-neighbor interactions tuned by the concentration $x$ show a magnetic phase diagram (T, $x$) in qualitative agreement with mean field predictions. Well-known examples are Ni$_{1-x}$Mn$_x$\cite{Abdul-Razzaq1984}, Au$_{1-x}$Fe$_x$\cite{Campbell1983}, Fe$_{1-x}$Al$_x$\cite{Motoya1983,Boeni1986} and Fe$_{1-x}$Cr$_x$\cite{Burke1983} crystalline alloys or amorphous Fe-based alloys \cite{Salamon1980,Birgeneau1978,Fernandez-Baca1990,Senoussi1988a,Senoussi1988b}. A large body of experimental and theoretical studies have revealed the peculiarities of their magnetic behavior. 

In this paper, we focus on vortex-like textures observed in the 1980s in the above systems, either in single crystal, polycrystal or amorphous form \cite{Hennion1986,Boeni1986,Lequien1987,Hennion1988}. They were detected under applied magnetic field in the mixed phases of ferromagnetic, weakly frustrated alloys ($x$ $\ll$ $x_{\rm C}$), using small angle neutron scattering (SANS), which provides a clear signature of these textures and reveals their typical size. Inside the vortices, the transverse spin components are frozen in the plane perpendicular to the applied field, and they are rotated over a finite length scale, yielding a maximum in the neutron scattering cross section versus the momentum transfer. In addition, the transverse spin freezing induces Dzyaloshinskii-Moriya (DM) anisotropy\cite{Campbell1986}, together with a chiral anomalous Hall effect\cite{Tatara2002,Pureur2004,Fabris2006}. Stimulated by these measurements, Monte Carlo (MC) simulations were performed in a 2d  lattice, showing similar vortex-like patterns\cite{Kawamura1991}. The knowledge of their spatial organization has however remained elusive.

In this context, it is worth recalling that ferromagnets may also host nanometric spin textures known as skyrmions (SKs). SKs form double-twist solitonic structures, offering many perspectives in spintronics and data storage\cite{Fert2013,Nagaosa2013}. As predicted by theory\cite{Okubo2012,Leonov2015,Lin2016,Rozsa2016,Hu2017}, some anisotropic ordered magnets with competing nearest neighbor (NN) and next nearest neighbor (NNN) exchange interactions 
may host localized SKs with versatile internal structure and smooth rotation of the magnetization. Different types of modulated phases such as hexagonal or square SK lattices have been predicted, yielding a very rich phase diagram \cite{Leonov2015}. 

The size of these "frustrated" SKs, of the order of a few lattice constants, is comparable to the typical vortex size in RSGs and much smaller than the size of chiral SKs stabilized by DM anisotropy in thin films or bulk state, which is usually above 10 nm \cite{Leonov2016,Leonov2016b}. Therefore, quenched disorder should affect frustrated SKs much more than their chiral counterparts, expected to undergo a collective pinning by disordered impurities without deep changes of their internal structure \cite{Hoshino2018}. Experimentally, large SK lattices were observed in non-centrosymmetric frustrated alloys with chemical disorder \cite{Tokunaga2015,Nayak2017}, showing magnetic anomalies similar to the RSG's. 
 Frustrated SKs have been suspected in very few systems so far, such as Gd$_{2}$PdSi$_{3}$ (Ref. \onlinecite{Kurumaji2018}). 

Remarkably, frustrated SKs reveal strong similarities with the vortex textures observed in RSGs. Our study attempts to clarify the subtle differences between these two types of topological defects. To that end, we report on new experiments performed on a weakly frustrated Ni$_{0.81}$Mn$_{0.19}$ single crystal, searching for a vortex lattice and aiming for a better characterization of these field-induced magnetic textures (Section \ref{sec:sectwo}). Our experiments are complemented by MC simulations with a minimal model, which clarifies the internal structure of the vortices and identifies their most relevant features (Section \ref{sec:secthree}). We discuss the origin of the vortex textures, and compare them with SKs, either chiral or frustrated, observed in bulk materials (Section \ref{sec:secfour}). 


\section{Vortex-like textures in a single crystalline reentrant spin glass}
\label{sec:sectwo}

\subsection{The Ni$_{1-x}$Mn$_x$ system and studied sample} 

%
\begin{figure*}[!ht]
		\centering
		\includegraphics[width=0.98\textwidth]{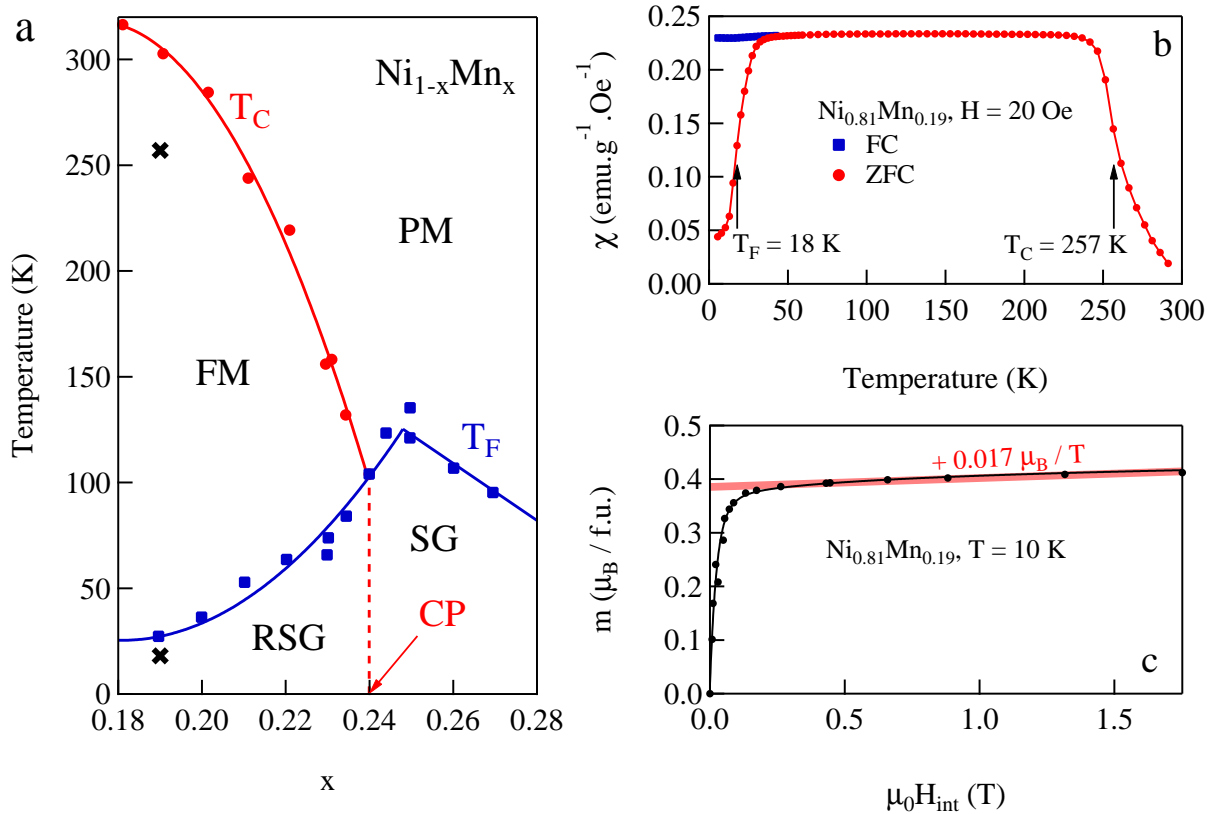}
		\caption{\label{fig:phasediagram}\textbf{(a)} Magnetic phase diagram of the disordered Ni$_{1-x}$Mn$_{x}$ system determined by bulk magnetic measurements and adapted from Ref. \onlinecite{Sommer1992}. A critical point (CP) occurs for $x \sim$ 0.24, see Ref. \onlinecite{Abdul-Razzaq1987}. \textbf{(b)} Static susceptibility $\chi$ of the Ni$_{0.81}$Mn$_{0.19}$ sample -initially zero field cooled down to 4.2\,K- recorded upon warming under a magnetic field of 20\,Oe (ZFC, red circles) then upon cooling under field (FC, blue squares). The characteristic freezing (T$_{\rm F} \sim 18\,$K) and Curie (T$_{\rm C} \sim 257\,$K) temperatures are obtained from the extrema of the temperature derivative of $\chi$ and reported in \textbf{(a)} (black crosses). \textbf{(c)} Field-dependence of the magnetization of Ni$_{0.81}$Mn$_{0.19}$ (reproduced from Ref. \onlinecite{Mirebeau1987}). After a quick rise, $m$ tends to saturate and further evolves with a small, yet finite, slope (red line).}
\end{figure*}

In Ni$_{1-x}$Mn$_x$ alloys, magnetic frustration arises from competing interactions between NN pairs, namely the AFM  Mn-Mn pairs and the FM Ni-Mn and Ni-Ni pairs\cite{Marcinkowski1961,Cable1974}. The NNN Mn-Mn pairs are FM. The multi-critical line between RSG and SG phases is located around $x_{\rm C} = 0.24$, close to the stoichiometric Ni$_3$Mn (see Refs. \onlinecite{Abdul-Razzaq1987,Sommer1992} and Fig. \ref{fig:phasediagram}a). Strikingly, the Ni$_3$Mn ordered superstructure of $L1_2$ type and space group $Pm\bar{3}m$ eliminates all NN Mn-Mn pairs. This  offers the possibility of tuning the magnetic order by controlling the number of such pairs through an appropriate heat treatment \cite{Yokoyama1976,Okazaki1995,Stanisz1989}. The fully ordered Ni$_3$Mn is a ferromagnet with a Curie temperature \TC\ $\sim$ 450\,K, whereas a disordered alloy of the same composition (space group $Fm\bar{3}m$) is a spin glass with a freezing temperature \TF\ $\sim$ 115\,K.  

Here, we study a Ni$_{0.81}$Mn$_{0.19}$ single crystal, already used for the neutron scattering experiments presented in Ref. \onlinecite{Hennion1988}. The single crystal form limits the distributions of magnetocrystalline anisotropies and demagnetizing fields within the sample, and provides the best playground to search for a vortex lattice. A thin rectangular plate was cut from the large crystal in a (110) plane for magnetic measurements. Both samples were heated at 900 $^{\circ}$C during 20 hours in a sealed quartz tube under vacuum, then quenched into an ice and water mixture to ensure maximal disorder\cite{Abdul-Razzaq1987}. They were stored in liquid nitrogen between experiments to prevent any further evolution of the short range order. 
 
Static magnetic susceptibility was measured versus temperature under a field H = 20 Oe in both field cooled (FC) and zero field cooled (ZFC) conditions, using a superconducting quantum interference device (SQUID). With decreasing temperature, the ZFC susceptibility strongly increases at the Curie temperature \TC\ = 257\,K,  shows a plateau over an extended temperature range as expected for weakly frustrated RSGs, and then decreases (Fig. \ref{fig:phasediagram}b). The freezing temperature \TF\ = 18\,K, defined similarly to \TC\ by the inflection point of the susceptibility versus temperature in the ZFC state, locates the onset of $\it{strong}$ magnetic irreversibilities. The ratio  T$_{\rm F}$/T$_{\rm C} \simeq 0.07$ characterizes the weak frustration of our sample. The canting temperature \TK\ $\sim$ 120\,K which situates between \TC\ and \TF\, locates much weaker irreversibilities related to transverse spin freezing. It was determined by previous neutron scattering experiments\cite{Lequien1987}. The three characteristic temperatures merge at the critical point. 

\subsection{Small-angle neutron scattering}

\begin{figure*}
		\centering
		\includegraphics[width=0.98\textwidth]{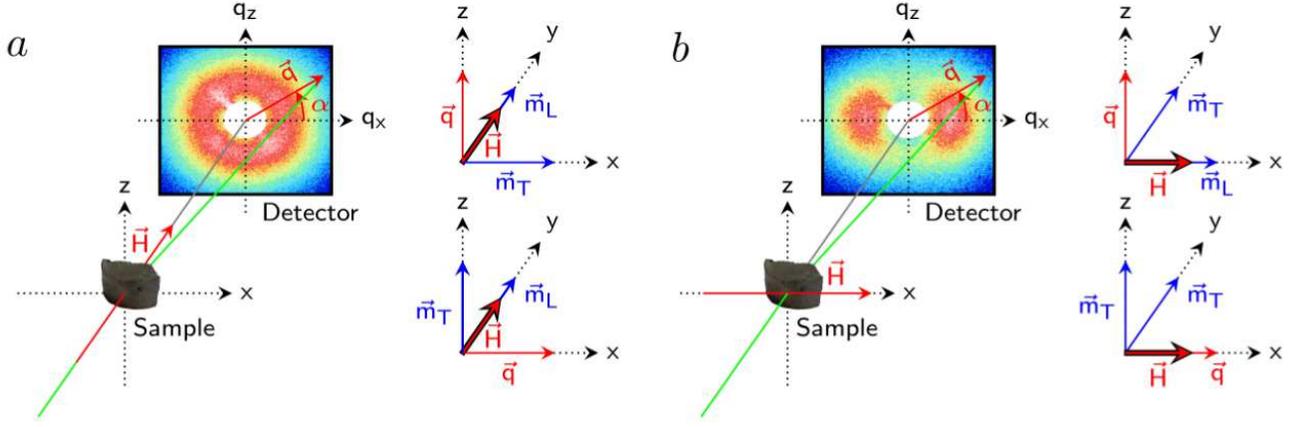}
		\caption{\label{fig:schemeconfig}Schematic configurations of the SANS experiments with the applied magnetic field $\mathbf{H}$ parallel to the neutron beam \textbf{(a)} and with $\mathbf{H}$ parallel to the detector plane \textbf{(b)}. Orientations of the transverse $\mathbf{m}_{\rm T}$ and longitudinal $\mathbf{m}_{\rm L}$ spin components are shown in each case which, combined to the selection rules of magnetic neutron scattering (Eq. \ref{eq:configb}), lead to the corresponding patterns (here recorded at T = 3 K and H = 2 T after zero-field cooling). In both configurations, the magnetic field is applied along the [110] (or equivalent) crystallographic direction.
}
\end{figure*}

SANS measurements were performed on the D33 instrument of the Institut Laue Langevin (ILL), using an incident neutron wavelength  $\lambda = 6~\text{\AA}$ and a sample to detector distance D = 2.8 m. Data were corrected for the detector efficiency and calibrated cross sections were obtained by taking the sample thickness and transmission, as well as the incident neutron flux, into account\cite{SM}. A magnetic field H  up to 2\,T was applied to the sample, in two configurations (see Fig. \ref{fig:schemeconfig}): \emph{a)} along the neutron beam, which defines the $y$ axis;  \emph{b)} along the $x$ axis perpendicular to the neutron beam, namely in a plane parallel to the detector $(x,z)$ plane.  Additional measurements were performed in configuration \emph{b)} on the PAXY spectrometer of the Laboratoire L\'eon Brillouin (LLB) under a magnetic field up to 8\,T for the same neutron wavelength and sample-to-detector distance. Fig. \ref{fig:schemeconfig} shows typical intensity maps recorded in the detector plane for the two configurations. The intensity is measured  at 3\,K  in the ZFC state under a magnetic field H = 2\,T, which almost saturates the sample magnetization (Fig. \ref{fig:phasediagram}c). 

In configuration \emph{a)}, the intensity distribution does not show any Bragg spot, rather a broad maximum at a finite momentum transfer. The intensity is isotropically distributed over a ring of scattering in the detector plane. The absence of any Bragg spots strikingly contrasts with the scattering patterns in SK lattices or superconducting flux line lattices observed in single crystal samples for the same experimental configuration\cite{Lynn1994,Muehlbauer2009}. 
It means that although the sample is single crystalline, the magnetic defects are organized in a random or liquid-like way.
As discussed below, this is due to the random occupation of the lattice sites and subsequent disorder of Mn-Mn NN AFM bonds. 

\begin{figure*}[!ht]
		\centering
		\includegraphics[width=0.98\textwidth]{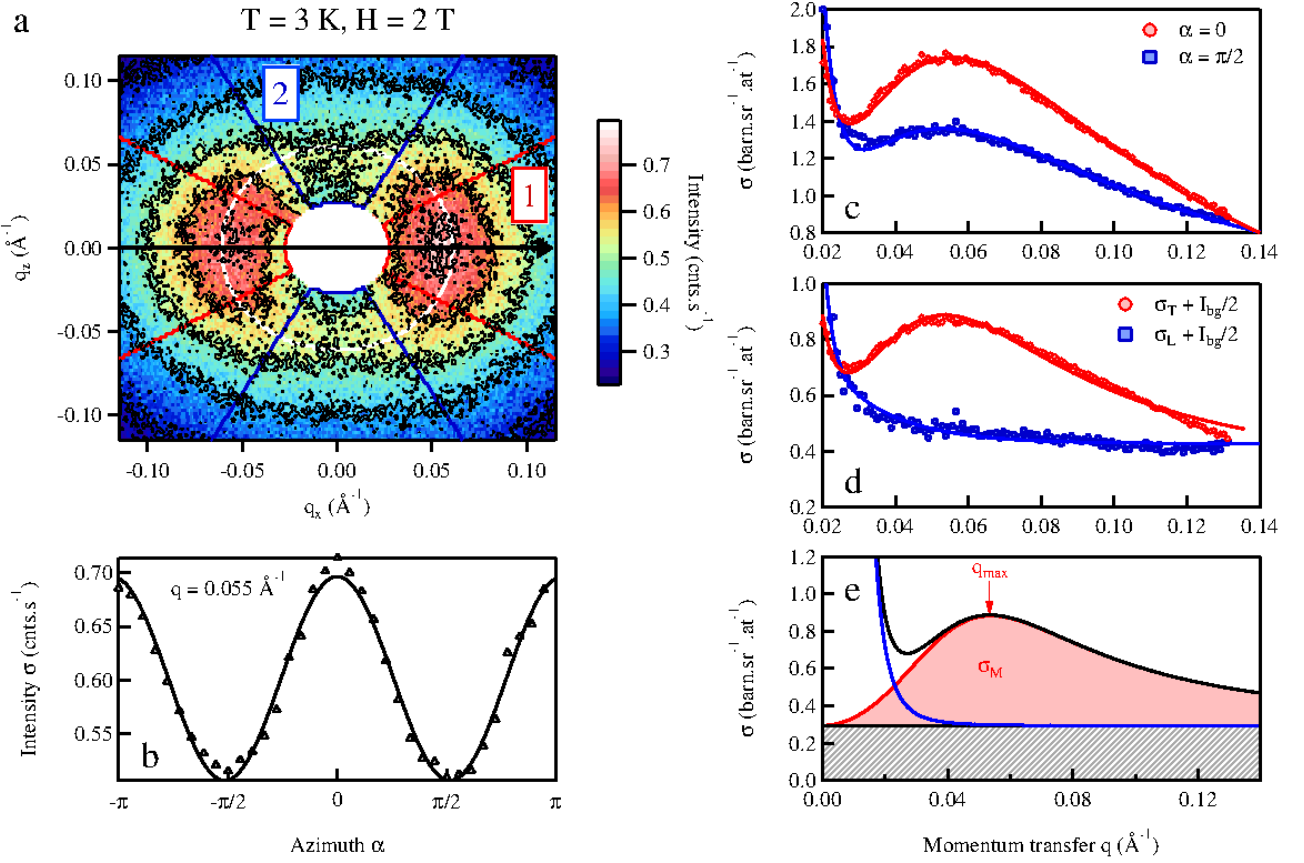}
		\caption{\label{fig:sans_setup}\textbf{(a)} Scattering map recorded at T = 3 K under a field H = 2 T applied perpendicular to the incident beam (black arrow). The two 60$^{\circ}$ angular ranges associated with $\mathbf{q}\, \| \,\mathbf{H}$ (red sector 1) and $\mathbf{q} \perp \mathbf{H}$ (blue sector 2) are also shown. \textbf{(b)} Angular dependence of the intensity at a momentum transfer $q = 0.055~\text{\AA}^{-1}$ ({\it i.e.} collected along the white circular trace in \textbf{(a)}). Solid line is a fit of Eq. \ref{eq:configb} to the data. \textbf{(c)} $q$-dependence of the scattering cross section for momentum transfers along (red circles) and perpendicular (blue squares) to the applied field. Solid lines with corresponding colors are fits of Eq. \ref{eq:sq_transverse} to the data. \textbf{(d)} $q$-dependence of the longitudinal ($\sigma_{\rm T}$) and transverse ($\sigma_{\rm L}$) magnetic cross sections, obtained by linear combinations of Eq. \ref{eq:configb} for $\alpha = 0$ ($\mathbf{q \, \| \, \mathbf{H}}$) and $\pi / 2$ ($\mathbf{q \perp \mathbf{H}}$). \textbf{(e)} Fit curve corresponding to the $\sigma_{\rm T}(q)$ data of panel \textbf{(d)} (black line) along with singled out peak function (red curve) and $q$-dependent background signal (blue curve). The $q$-independent background contribution is also shown (hatched area, see text and Ref. \onlinecite{SM}).}
\end{figure*}

In configuration \emph{b)}, one observes a similar pattern, but the intensity is now modulated according to the orientation of the momentum transfer with respect to the applied field, and is strongly enhanced in the direction $\mathbf{q} \, \| \, \mathbf{H}$. This modulation comes from the selection rule for magnetic neutron scattering, which impose that only the spin components perpendicular to the scattering vector $\mathbf{q}$ contribute to the magnetic cross-section. As schematically explained in Fig. \ref{fig:schemeconfig} the dominant contribution to the scattering in this configuration arises from spin components $\mathbf{m}_{\rm T}$ transverse to the magnetic field. In the following analysis, we focus on this configuration, which allows us to better characterize the spin textures. The intensity maps in configuration \emph{b)} can be described as

\begin{eqnarray}
	\nonumber
	\sigma(q,\alpha) &=& \sigma_{\rm L}(q) \cdot \sin^2\alpha \\ 
	&+& \sigma_{\rm T}(q) \cdot (1+\cos^2\alpha)+I_{\rm bg}(q) \quad ,
	\label{eq:configb}
\end{eqnarray} 

\noindent
where $\alpha$ is the angle $(\mathbf{q},\mathbf{H})$, $\sigma_{\rm L}(q)$ and $\sigma_{\rm T}(q)$ are the magnetic scattering cross sections related to correlations between transverse and longitudinal spin components, respectively. $I_{bg}$(q) is an isotropic background which consists of a low-q contribution from crystal inhomogeneities and a constant term which can be calculated exactly, and which is in excellent agreement with experiment (see details in Ref. \onlinecite{SM}).
Noticing that Eq. \ref{eq:configb} fits the angular dependence of the intensity, we average the scattering map within two angular sectors of 60$^{\circ}$: sector 1 for $\mathbf{q} \, \| \, \mathbf{H}$ ($\alpha$ = 0$^{\circ}$) and sector 2 for $\mathbf{q} \perp \mathbf{H}$ ($\alpha$ = 90$^{\circ}$) (see Fig. \ref{fig:sans_setup} a,b and c). We then combine the intensities from the two sectors to separate the contributions from  the transverse and the longitudinal spin components (Fig. \ref{fig:sans_setup}d). 

\begin{figure*}[!ht]
		\centering
		\includegraphics[width=0.98\textwidth]{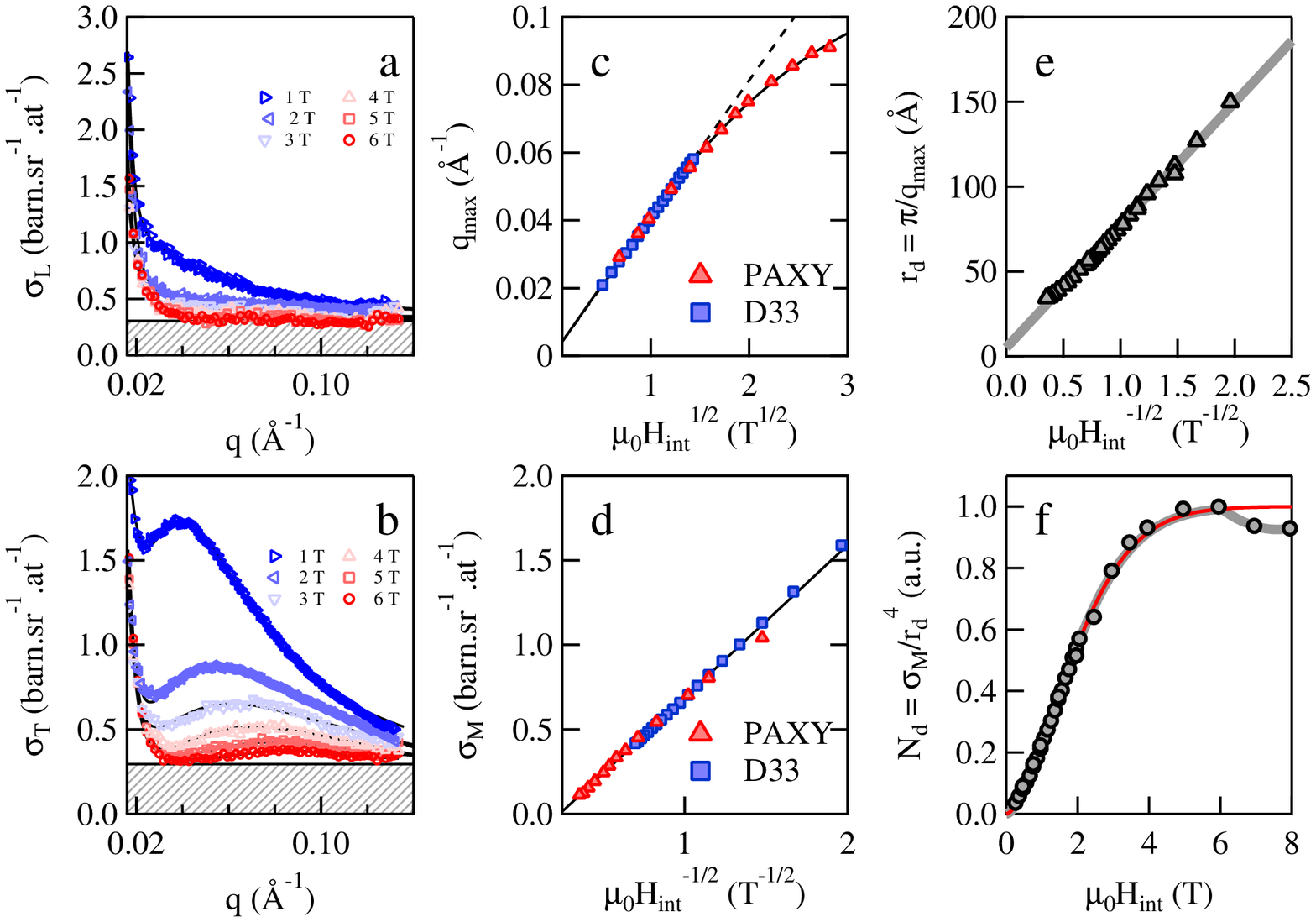}
		\caption{\label{fig:scaling}Evolution of the longitudinal \textbf{(a)} and transverse \textbf{(b)} magnetic neutron scattering cross sections as a function of magnetic field. In both panels, the hatched areas represent the q-independent background contribution (see text and Ref. \onlinecite{SM}). Field-dependence of the peak position $q_{\rm max}$ \textbf{(c)} and integrated scattering intensity $\sigma_{\rm M}$ \textbf{(d)} as obtained from a fit of Eq. \ref{eq:sq_transverse} to the SANS data from D33 (blue squares) and PAXY (red triangles). \textbf{(e)} Field-dependence of the defect size $r_{\rm d}$, computed using data of panel \textbf{(c)}. \textbf{(f)} Field-dependence of the number of scattering centers $N_{\rm d}$, as seen by SANS. The red line is a fit of Eq. \ref{eq:stretched_exp} to the data, in the $0 \leq \mu_{\rm 0}H_{\rm int} \leq 6 \,\text{T}$-range.}
\end{figure*}

As a key result, the intensity from the transverse spin components $\sigma_{\rm T}(q)$ shows a clear maximum  in $q$, which arises from the vortex-like textures.  As shown below, the FM correlated transverse spin components rotate over a finite length scale to compensate the transverse magnetization, yielding negligible intensity at $q = 0$ and a maximum related to the vortex size. When the field increases, the maximum intensity decreases and its position moves towards high q values (Fig. \ref{fig:scaling}b). A signal from the transverse spin components is observed up to the highest field of 8\,T. On the other hand, the intensity from the longitudinal spin components $\sigma_{\rm L}(q)$ shows no well-defined maximum at $q \neq 0$ (Fig. \ref{fig:scaling}a). Above 2\,T, it becomes very small and difficult to separate from the background contribution\cite{SM}. 
 
In a first step, the transverse cross section was fitted by the phenomenological expression    

\begin{eqnarray}
	\nonumber
	\sigma_{T} (q) &=& \frac{\sigma_{\rm M} \, \kappa \, q}{2\pi q_{\rm 0}} \cdot \left(\frac{1}{\kappa^2+\left(q-q_{\rm 0}\right)^2}-\frac{1}{\kappa^2+\left(q+q_{\rm 0}\right)^2}\right)\\
&+& \frac{I_{\rm bg}(q)}{2} \quad ,
	\label{eq:sq_transverse}
\end{eqnarray}
 where the first term accounts for the observed peak in the scattering cross section while the second one is related to the background. 
 From Eq. \ref{eq:sq_transverse}, one can extract the peak position $q_{\rm max} = \sqrt{q_{\rm 0}^{2}+\kappa^{2}}$ and the integrated cross section $\sigma_{\rm M}$. As shown in Fig. \ref{fig:scaling}c-f, these quantities vary continuously with the magnetic field.     


To interpret these results, we take into account the liquid-like order of the defects in analogy with chemical inhomogeneities. Having fitted and subtracted the background term, we express the scattering cross section as

\begin{eqnarray}
	\nonumber
  \sigma_{T} (q) &=&a\, \Delta \rho_{\rm mag}^{2} \, N_{\rm d} \, V_{\rm d}^{2}\\
	&\times& \left\{\langle F_{\rm T}^{2}(q)\rangle - \langle F_{T}(q) \rangle^{2} \cdot \left[ 1 - S_{\rm int}(q) \right] \right\} \quad , 
	\label{eq:iq_transverse}
\end{eqnarray}   

\noindent
where $F_{\rm T}(q)$ is the normalized form factor of the defects, associated with transverse spin components, and $S_{\rm int}(q)$ is an interference function which takes into account the local correlations between two defects. In Eq. \ref{eq:iq_transverse}, $\langle\rangle$ denotes the statistical average over the sample.
 $\Delta \rho_{\rm mag} \simeq \left|\mathbf{m}_{\rm T}\right|$ is the magnetic contrast between a vortex (where $\left|\mathbf{m}_{\rm T}\right| \neq 0$) and the surrounding ferromagnetic region (where $\left|\mathbf{m}_{\rm T}\right| \rightarrow 0$). 
$N_{\rm d}$  and $V_{\rm d}$  are respectively the number of vortices and their volume, and $a$ is a constant.

In the following, we neglect the local magnetic interaction between defects. This assumption of independent objects is justified for a weakly frustrated system where the vortex centers are randomly distributed and located far away from each other ({\it i.e.} $S_{\rm int}(q) = 1$ in Eq. \ref{eq:iq_transverse}). This assumption also holds for a system with concentrated defects, taking into account the specific form factor of the magnetic vortices and the random orientation of the transverse spin components from one vortex to another ({\it i.e.} $\langle F_{T}(q) \rangle =0$ in Eq. \ref{eq:iq_transverse}). It is confirmed by analytical calculations of model form factors \cite{SM} and by MC simulations reported in Section \ref{sec:secthree}.  

For independent defects, the $q$-dependence of the neutron intensity reduces to that of the average squared form factor, and the position $q_{\rm max}$ of the intensity maximum is inversely proportional to the typical size of the vortices. The integrated intensity $\sigma_{\rm M}$ is proportional to $\Delta \rho_{\rm mag}^{2} \, N_{\rm d} \, V_{\rm d}^{2}$, according to Eq. \ref{eq:iq_transverse}. As a toy model, we have considered regular vortices of radius $r_{\rm d}$ having an antiferromagnetic core\cite{SM}. The squared form factor averaged over all orientations for the transverse components has a non symmetric line shape recalling the experimental one, with a maximum at $q_{\rm max} = \pi / r_{\rm d}$.

Therefore, taking into account corrections for the demagnetization factor, the field dependence of $q_{\rm max}$ reflects the decrease of the vortex typical radius  $r_{\rm d} = \pi / q_{\rm max}$ with increasing field\cite{SM}. Over the explored field range, $r_{\rm d}$ obeys the simple relation $r_{\rm d} \propto H^{-1/2}$ (Fig. \ref{fig:scaling}e). The corresponding variation of $\sigma_{\rm M} \propto H^{-1/2}$ suggests that the evolution of the defect shape versus the magnetic field occurs in a self similar way, yielding scaling laws for the position, width and intensity of the magnetic signal (Fig. \ref{fig:scaling}b). Such laws are actually quite general and, for instance, govern the evolution of the cluster size with annealing time in metallic alloys which tend to segregate when they are quenched in the region of spinodal decomposition \cite{Hennion1982}.

Using Eq. \ref{eq:iq_transverse}, one can also infer the field-dependence of the number of defects ({\it i.e.} scattering centers) seen by SANS from  the quantity $\sigma_{\rm M} / V_{\rm d}^{2}$. For this purpose we assume thin cylindrical defects, and consider an experimental field range $H \ll J$ where the magnetic contrast (or amplitude of the transverse spin component) is roughly field-independent.  We obtain $V_{\rm d} \simeq r_{\rm d}^{2}$, thus $N_{\rm d} \simeq \sigma_{\rm M} / r_{\rm d}^{4}$. As shown in Fig. \ref{fig:scaling}f, $N_{\rm d}$ \emph{increases} with increasing field and saturates at a finite field of $\simeq$ 6 T. This variation is described by a stretched exponential

\begin{equation}
	\label{eq:stretched_exp}
	N_{\rm d} = 1 - \exp \left(-\frac{H}{H_{\rm C}}\right)^{\nu} 
\end{equation}

\noindent
 with $H_{\rm C} = 2.29(3)\,$T and $\nu = 1.64(4)$.  
This result can be understood as follows. At low fields, vortices are large enough to involve several AFM bonds. Upon an increase in field, they progressively shrink while remaining centred on isolated AFM first neighbor Mn-Mn pairs\cite{SM}, the number of which is fixed by the Mn concentration and heat treatment. Consequently, the number of individual defects $N_{\rm d}$ seen by SANS will \emph{increase}. At higher fields, however, $N_{\rm d}$ should decrease until all defects have collapsed for fields strong enough to overcome the typical AFM exchange interaction. We indeed observe a slight decrease of $N_{\rm d}$ for $\mu_{\rm 0}H_{\rm int} \gtrsim $ 6 T. However we note that the field corresponding to the exchange interaction is of the order of several 100\,T and is thus well-beyond our experimental range. In turn, this regime can be conveniently explored numerically. This point is addressed in the next section, where we propose a way to verify the above scenario and extend the exploration of the vortex-like textures properties towards arbitrarily large magnetic fields.  

\section{Monte Carlo simulations}
\label{sec:secthree}

\begin{figure*}[!ht]
         \centering
         \includegraphics[width=0.98\textwidth]{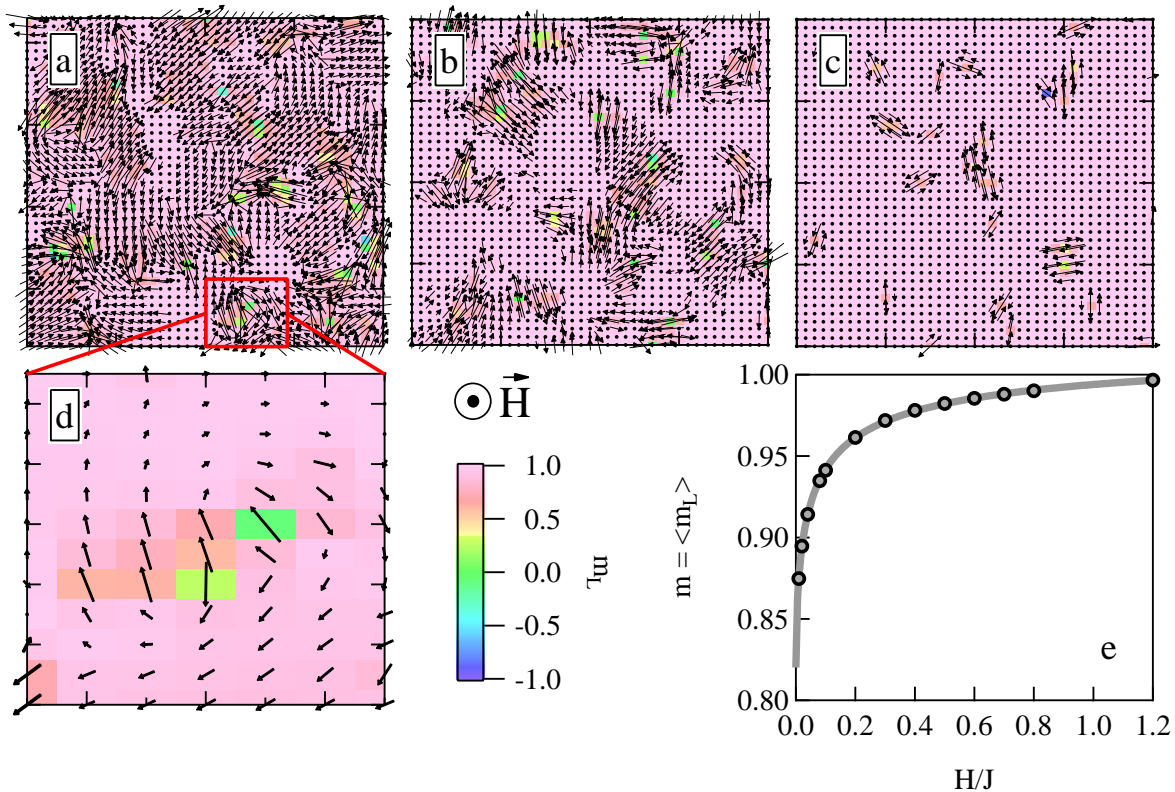}
         \caption{\label{fig:mc_results_rspace} Outcome of the Monte Carlo simulations for an antiferromagnetic bond concentration of 5 \%. 
\textbf{(a-c)} Simulated spin maps for H/J = 0.01, 0.1 and 0.8, respectively, where J is the first neighbor exchange term. We display portions of 40x40 spins for clarity. The longitudinal magnetization $m_{\rm L}$ ({\it i.e.} parallel to the applied field) is color-coded while the arrows represent the transverse component $m_{\rm T}$. \textbf{(d)} Example of a vortex-like chiral spin circulation stabilized at H = 0.01 J. \textbf{(e)} Field-dependence of the bulk magnetization $m$, defined as the average value of the longitudinal magnetization per spin $\langle m_{\rm L} \rangle$.} 
\end{figure*}

Numerical studies of the reentrance phenomena and magnetic structures of reentrant spin glasses trace back to the pioneering work of Kawamura and Tanemura\cite{Kawamura1986,Kawamura1991}. They showed that a minimal model is able to reproduce the main characteristics of the magnetic textures observed in RSG's. Following their approach, we first performed MC simulations on 2d matrices containing $160 \times 160$ Heisenberg spins placed on a square lattice. While the main interaction is assumed to be FM ($J = 1$), a certain fraction $c$ of the bonds is turned into AFM ($J = -1$). Using a spin quench algorithm, the system's ground state is found where vortex-like defects appear as metastable configurations ({\it i.e.} with energies slightly higher than those of the bulk FM state). For the studied concentrations $c = 5\,\text{and}\, 20 \, \%$, individual defects (similar to vortices or pairs of vortices) are evidenced, all of them being centred around the randomly distributed AFM NN pairs (see Fig. \ref{fig:mc_results_rspace} for the 5\% case and Ref. \onlinecite{SM} for further details). 

In all cases, the average topological charge is $Q = 0$ but individual objects locally display a finite $Q$, being in some cases as large as 0.3 ({\it i.e.} similar values as those found for certain types of frustrated SKs\cite{Leonov2015}). The origin of the non-integer charge is clarified by considering the relatively small size of the defects as well as their irregular shapes and distorted magnetization profiles, related to the ill-defined boundaries between vortices and the ambient FM medium. In other words, the vortex-like textures stabilized under field in RSGs feature both senses of the vector chirality, resulting in a smaller topological charge than in frustrated or chiral SKs (for which $Q = \pm 1$).
Extending the MC simulations to a 3d spin matrix, it appears that the vortex-like textures keep their anisotropic shape (oblate along the field direction) in the 3d lattice and can thus be dubbed as "croutons".

As shown qualitatively on the maps displayed in Figs. \ref{fig:mc_results_rspace}a-c, the average number of defects decreases with increasing field $H$ while spins are progressively aligned along its direction. The computed magnetization $m$ (Figs. \ref{fig:mc_results_rspace}e) increases as the number of vortices decreases, showing a quasi plateau with finite slope versus the ratio $H/J$. At very high fields, of magnitude comparable to the exchange constant $J$, a prediction of the MC modeling is that the vortices should collapse individually, yielding microscopic plateaus of $m$, the amplitude of which is probably too small to be experimentally observed.


 
\begin{figure*}[!ht]
         \centering
         \includegraphics[width=0.98\textwidth]{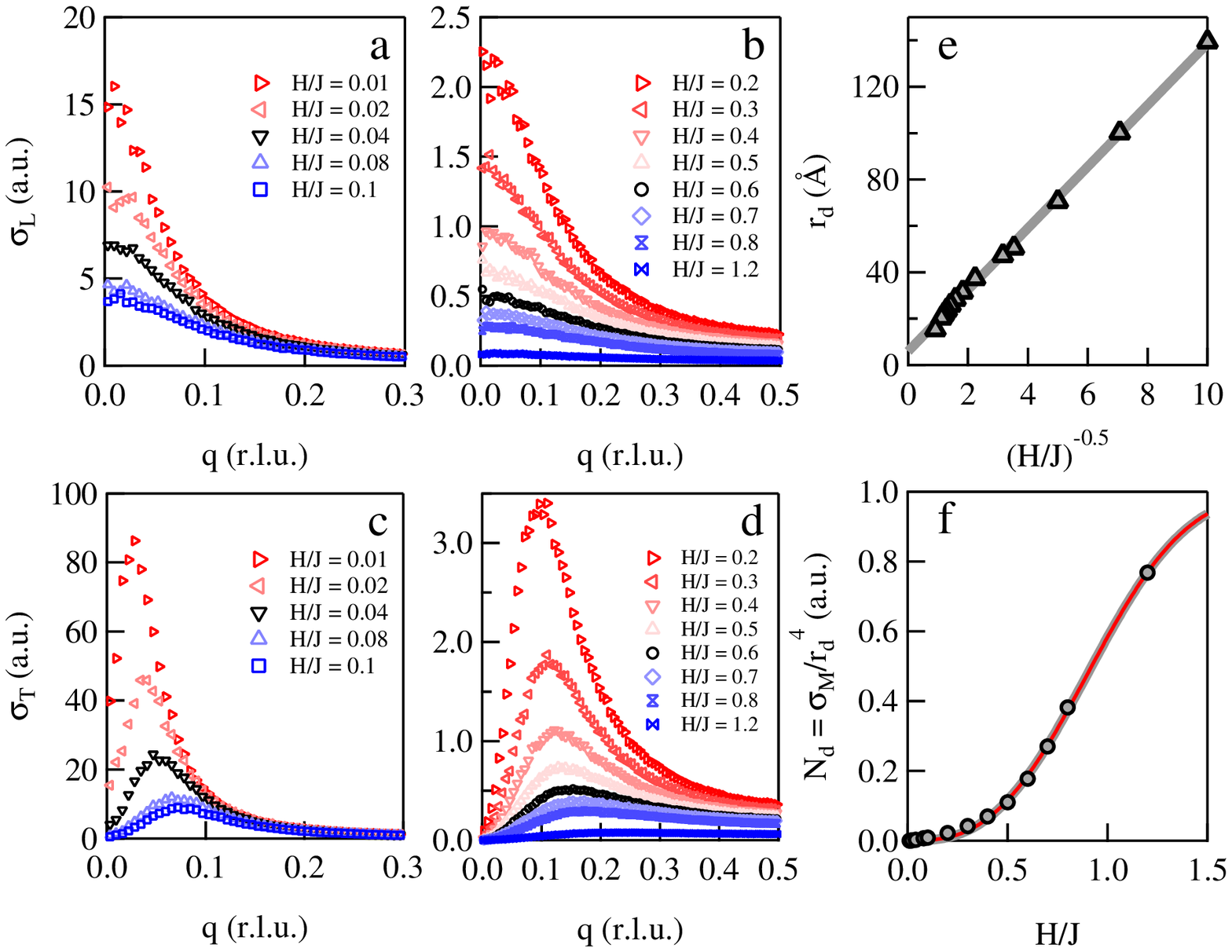}
         \caption{\label{fig:mc_results_qspace}Fourier analysis of the Monte Carlo results of Fig. \ref{fig:mc_results_rspace}. \textbf{(a-d)} Computed longitudinal \textbf{(a,b)} and transverse \textbf{(c,d)} form factors, obtained by Fourier transforming the calculated spin maps for different field values. The momentum $q$ is expressed in reciprocal lattice units (r.l.u.). \textbf{(e)} Field-dependence of the average defect size $r_{\rm d} = a / q_{\rm max}$, where $q_{\rm max}$ is the position of the peak in $\sigma_{\rm T}$ and $a = 3.586$ \AA\, is the lattice constant of Ni$_{0.81}$Mn$_{0.19}$. \textbf{(f)} Field-dependence of the apparent number of defects $N_{\rm d}$. The red line is a fit of Eq. \ref{eq:stretched_exp} to the data, with an inflection point at $H_{\rm C} = 1.05(5) \, J$ and a stretching exponent $\nu = 2.8(1)$.} 
\end{figure*}

In order to compare these results with the SANS experiments of Section \ref{sec:sectwo}, we have computed the Fourier transforms of the longitudinal and transverse spin components. The longitudinal cross section $\sigma_{\rm L}$ decreases monotonically with increasing $q$ (Fig. \ref{fig:mc_results_qspace}a,b) whereas the transverse cross section $\sigma_{\rm T}$  shows a broad asymmetric peak (Fig. \ref{fig:mc_results_qspace}c,d).  Both quantities become almost $q$-independent at large $q$ values. When the field increases, the magnitude of the two simulated cross sections decreases, and a fit of Eq. \ref{eq:sq_transverse} to the simulated $\sigma_{\rm T}$-curves shows that the position of the maximum $q_{\rm max}$ moves towards larger values, whereas its integrated intensity $\sigma_{\rm M}$ decreases. 
This evolution reflects a decrease of the vortex size $r_{\rm d}$ with increasing $H$ according to a scaling law (Fig. \ref{fig:mc_results_qspace}e) and an apparent increase of the number of vortices $N_{\rm d}$ following  Eq. \ref{eq:stretched_exp} with fit parameters $H_{\rm C} = 1.05(5) \,J$ and $\nu = 2.8(1)$ (Fig. \ref{fig:mc_results_qspace}f). Similar to the experimental case, $N_{\rm d}$ is defined as $N_{\rm d} \simeq \sigma_{\rm M} / r_{\rm d}^{4}$, where $r_{\rm d} = a / q_{\rm max}$ with $a$ the lattice constant of Ni$_{0.81}$Mn$_{0.19}$. As discussed below, these results show that a minimal model is able to capture the essential features of the observed textures.


\section{Discussion}
\label{sec:secfour}

\subsection{Spin textures in a reentrant spin glass: the "crouton" picture}

 The MC simulations presented above strongly reflect the experimental observations, as shown by:
\begin{enumerate}
	\item The shape of the magnetization curve with a finite slope at large fields (compare Fig. \ref{fig:phasediagram}c and \ref{fig:mc_results_rspace}e),	
	\item The existence of defects over which the transverse magnetization is self-compensated, yielding a peak of $\sigma_{T}$ at a finite $q$-value. The asymmetric q-dependence of $\sigma_{T}$ is also reproduced, suggesting similar internal structures of the defects (compare Fig. \ref{fig:scaling}b and \ref{fig:mc_results_qspace}c,d),
	\item The persistence of inhomogeneities of the magnetization at the scale of the vortex size, deep inside the RSG phase, as shown by the finite longitudinal cross section $\sigma_{L}$ centered around $q = 0$ (compare Fig. \ref{fig:scaling}a and \ref{fig:mc_results_qspace}a,b),
	\item The field-dependence of the defect size $r_{\rm d}$ (obtained from the $q$-position of the peak in $\sigma_{T}$), obeying scaling laws $r_{\rm d} \propto H^{-\beta}$ with the same exponent $\beta = 0.5$ (compare Fig. \ref{fig:scaling}e and \ref{fig:mc_results_qspace}e), 
	\item The field-dependence of the number of individual defects $N_{\rm d}$, increasing as a function of field following the phenomelogical Eq. \ref{eq:stretched_exp}, before reaching saturation (compare Fig. \ref{fig:scaling}f and \ref{fig:mc_results_qspace}f), 
	\item The robustness of the defects, surviving up to very large fields as compared with usual magnetic SKs (compare Fig. \ref{fig:scaling}f and \ref{fig:mc_results_qspace}f, and see Ref. \onlinecite{SM} for a detailed discussion).
\end{enumerate}
 
Therefore, the simulations strongly support a description of the magnetic defects observed in Ni$_{0.81}$Mn$_{0.19}$ as "crouton-like" defects, induced by AFM Mn-Mn first neighbor pairs, where the transverse spin components are ferromagnetically correlated and rotate to compensate the transverse magnetization. Their magnitude decreases from the vortex center to the surroundings to accommodate the average ferromagnetic medium. As discussed below, such defect shape is compatible with the interactions generally considered for the RSGs, although other defect textures could be in principle compatible with the experiment. 

%
%
%
\begin{figure*}[!ht]
\includegraphics[width=0.9\textwidth]{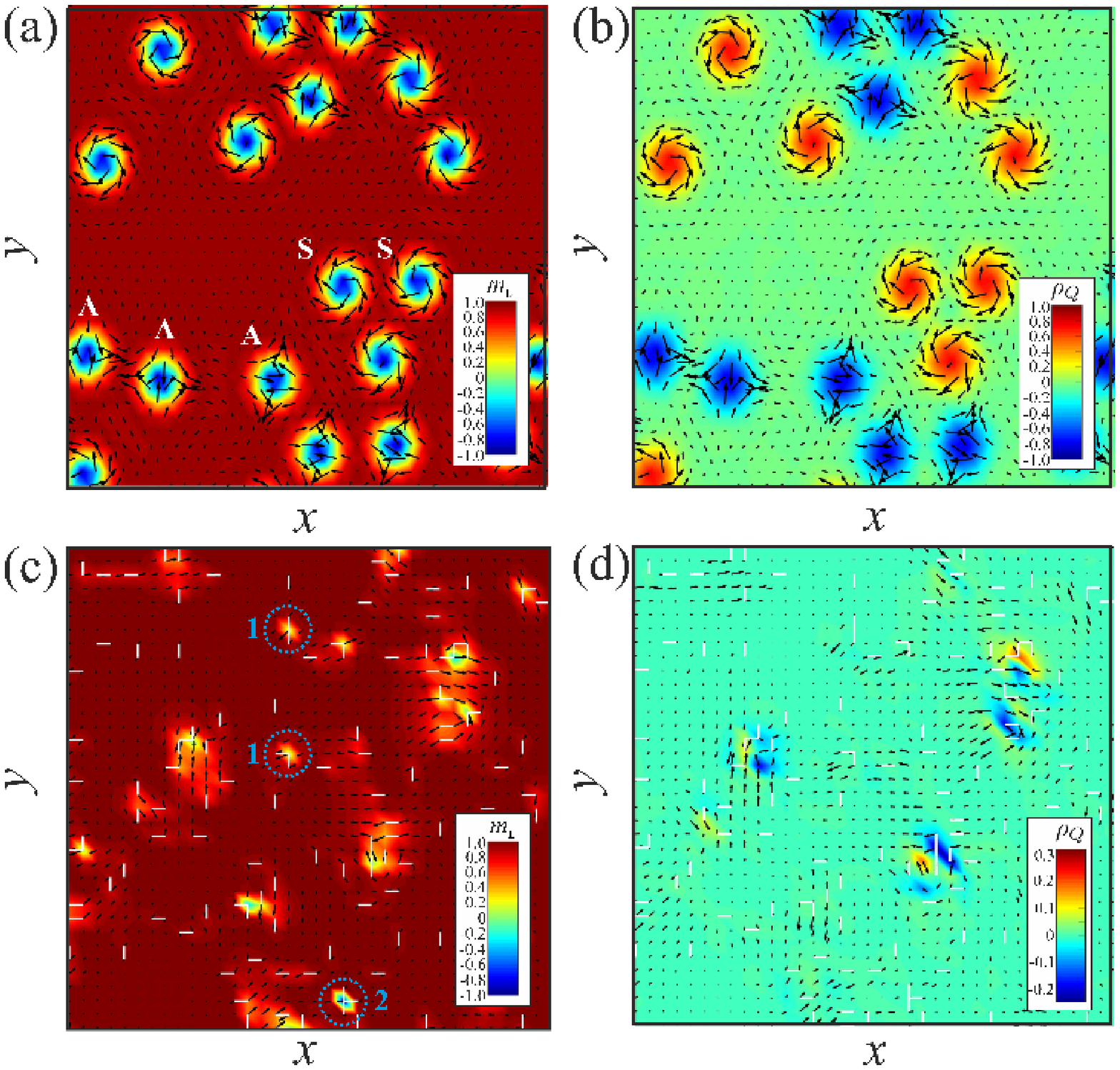}
\caption{
\label{skyrmions} \textbf{(a)}, \textbf{(b)} Skyrmions (S) and antiskyrmions (A) in a frustrated ferromagnet with competing NN  and NNN  exchange interactions $ J_1$ and $J_2$ under a field $h$, with $J_2/J_1=0.5,\,h=0.4$. Typical metastable states featuring clusters of skyrmions and antiskyrmions are obtained by relaxing the magnetic configuration with a random initial spin configuration. Color plots of $m_L$-components \textbf{(a)} and the topological charge density $\rho_Q$ \textbf{(b)} reflect the smooth rotation of the magnetization within the skyrmion cores. 
\textbf{(c)}, \textbf{(d)} Vortex-like defects induced by interaction disorder as described by a model  with random interaction $J_{ij}$ between  $i,j$ sites under a field $h$. The  $J_{ij}$ are independent random variables taking the values +1 and -1 with probabilities $1-c$ and $c$ respectively, with $c=0.05$ and $h=0.09$. Color plots of $m_L$-components \textbf{(c)} exhibit various types of spatially localized objects with the balanced topological charge density \textbf{(d)}. Blue numbered circles show vortices residing on two or three AFM bonds with distinct magnetization distributions. See  Ref. \onlinecite{SM} for details.
}
\end{figure*} 
The main difference between the experiment and the MC simulations is the field value at which the number of individual defects $N_{\rm d}$ starts saturating ($H \ll J$ and $\sim J$, respectively). This suggests that an accurate determination of their stability range requires a more complex modeling, which is well-beyond the scope of the present work. Indeed, the experimental situation is complicated, involving different moments on Ni and Mn ions, 3 types of interactions, a 3d lattice with high connectivity, a high concentration of magnetic species, and an atomic short-range order (SRO).  Therefore many different local environments and moment values exist in the experimental system. Comparatively, the simulations are based on a very simple case, namely a 2d square lattice with a random distribution of AFM bonds involving a single exchange constant. 

Despite these differences,  we stress that the agreement between both approaches is surprisingly good. Let us outline several reasons for that. Firstly, the mean field description, which identifies longitudinal and transverse spin components with different behaviors, is valid, as expected for weak frustration. The present sample behaves as a weakly frustrated ferromagnet (the ratio T$_{\rm F}/$T$_{\rm C} \simeq 0.07$ can be associated to an effective concentration of AFM bonds of $\simeq$ 0.07 in mean field approximation\cite{SM}), although the concentration of first neighbor isolated Mn-Mn pairs is relatively high (in the 0.2-0.4 range depending on the amount of SRO). Experiments varying the degree of frustration through Mn content or heat treatment could check the validity of this description when approaching the critical point which separates RSG and SG phases.

Secondly, both methods involve \emph{a statistical average of different types of defects} which do not interact with each other, but all have a typical size governed by general stability equations. This typical size is dictated by the competition between ferromagnetic exchange ($E = J k^2$) and Zeeman energy, and it is expected to vary as $k^{-1} \propto (J/H)^{0.5}$, hence $r_{\rm d} \propto H^{-0.5}$, as observed experimentally and in the simulation. Such a general law also controls the extension of Bloch walls\cite{Rado1982} or soliton defects\cite{Steiner1983} among others. 
 
Our findings also suggest that the 2d lattice provides a relevant description of the real case due to the peculiar crouton shape, with much larger extension in the transverse plane than along the field axis. In 2d-XY antiferromagnets, spontaneous vortices are stabilized and undergo a Kozterlitz -Thouless transition with temperature, involving spontaneous symmetry breaking at a local scale \citep{Kosterlitz1973,Kosterlitz2017,Villain2017}. The reentrant transitions have a different nature, but they also involve peculiar symmetry breaking  below \TK\ and \TF, associated to the Gabay-Toulouse and de Almeida-Thouless lines respectively\cite{Gabay1981}. As a major consequence, the transverse spin freezing and emergence of vortices strongly impact the spin excitations. A softening of the spin wave stiffness\cite{Hennion1982,Hennion1988}occurs below \TK, recalling the anomalous sound velocity in glasses\cite{Black1977,Michel1987} and the spin wave softening in quasi 2d frustrated antiferromagnets\cite{Aristov1990}. It is followed by a further hardening of the spin waves below \TF.
\subsection{Vortex-like textures and skyrmions}
Among the various classes of spin textures\cite{Toulouse1976}, those studied here show clear differences with the Bloch-type skyrmions observed in bulk chiral ferromagnets, which are primarily induced by DM anisotropy in non centrosymmetric lattices. Both occur in an average ferromagnetic medium, but the vortex-like textures probed in this study are stabilized at low temperatures, do not form a magnetic lattice and can exist for any crystal symmetry or even in amorphous compounds. This is because their primary origin is the competition of (symmetric) exchange interactions combined with site disorder, rather than antisymmetric exchange. In frustrated systems, the role of the latter, yielding DM anisotropy of chiral nature, has been investigated both theoretically \cite{Soukoulis1983,Gingras1994} and experimentally\cite{Campbell1986}. DM interactions explain the macroscopic irreversibilities in spin glasses and RSGs\cite{Senoussi1983}, torque measurements and paramagnetic resonance. Under field cooling conditions, they induce an additional magnetic field of unidirectional nature, which explains the slight decrease of the vortex size in NiMn when the sample is field-cooled\cite{SM,Mirebeau1988}. However they play a minor role in the stabilization of the vortex state, as exemplified by the MC results which describe a bare Heisenberg system. Experimentally we point out that across the critical concentration, the vortices disappear in the true spin glass phase, while the DM anisotropy hardly changes \cite{Martin2018}.
 
The dissimilarities between vortices and frustrated SKs are more subtle but can be understood by a comparative analysis of their internal structure  and topological charge, as deduced from MC simulations. This comparison is made in details in Ref. \onlinecite{SM}, and its main results are shown in  Fig. \ref{skyrmions}. Essentially, frustrated SKs are predicted in \emph{ordered} anisotropic magnets with competing interactions and inversion symmetry and do not require antisymmetric exchange\cite{Okubo2012,Leonov2015,Lin2016}. Their center correspond to a magnetic moment $\mathbf{m}$ antiparallel to the applied field ($m_{\rm L} < 0$), which gradually rotates towards the aligned state at the boundary ($m_{\rm L} > 0$). Therefore, they individually possess a large topological charge (+1 or -1) and can form densely packed clusters. These features contrast with the vortex-like textures studied in this work, pinned by locally disordered AFM bonds and unable to form extended ordered phases. Since $m_{\rm L}$ can alternate in sign in the vortex core (as controlled by their internal bond structure), the vortices neither bear smooth rotation of $\mathbf{m}$ nor select a preferred helicity, and their absolute topological charge density is smaller than unity within isolated defects. 

Although they cannot form ordered phases, the vortices studied in the present work may form a liquid-like order in the limit of small applied magnetic fields and large concentration of AFM bonds. Finally, both vortices and frustrated SKs remain metastable solutions, found at zero or finite temperatures. They are both endowed with a remarkable robustness against the collapse towards the field-induced FM state and do not require well-defined lattice structures to appear.

\section{Outlook and Conclusion}
   
Our study suggests that in the general search for SK-hosting systems, the role of disorder should be investigated more intensively. Its main consequences are expected in the low energy dynamics, associated with glassy states which occur both in average ordered or disordered media. Theoretically, the glassy behavior is related to metastable states with hierarchical structure in the ground state manifold. It can be analyzed in terms of replica field theory, initially developed for spin glasses \cite{Sherrington1975}, then extended to RSGs \cite{Gabay1981}, vortex lattices in superconductors (the Bragg glass phases \cite{Giamarchi1995}) and very recently to the skyrmion glass phase \cite{Hoshino2018}. The example of Co-Zn-Mn alloys \cite{Tokunaga2015} where SK lattices are observed at 300\,K and above, is an interesting playground to study such aspects in detail. There, site inversions should lead to local frustration effects and possibly explain the metastable textures observed experimentally. 

Our work constitutes an experimental illustration of the importance of frustration and disorder for the emergence of localized spin textures in condensed matter. We suggest a simple mechanism for tuning their properties (density, size) by different parameters such as the magnetic field, heat treatment or concentration. This could open a promising route towards the engineering of bulk systems with well defined sizes and density ranges, for instance the design of vortices by a controllable distribution of bonds. Moreover, while the observed vortices cannot be moved since they are bound to the Mn-Mn pairs, their interaction with electric currents\cite{Prychynenko2018,Bourianoff2018} and spin waves\cite{Continentino1983,Korenblit1986,Shender1991} is non trivial. In both cases, the class of frustrated ferromagnets studied in this paper might offer novel ways to encode complex information into electron and heat pulses.

\section*{Acknowledgements}

We thank P. Bonville for the SQUID measurements, M. Bonnaud for technical assistance on the D33 spectrometer and S. Gautrot and V. Thevenot for their help in setting the experiment on PAXY. A.O.L. acknowledges JSPS Core-to-Core Program, Advanced Research Networks (Japan) and JSPS Grant-in-Aid for Research Activity Start-up 17H06889, and thanks Ulrike Nitzsche for technical assistance.


\pagebreak
\appendix

\begin{center}
{\it \Large Supplementary material}
\end{center}

In this supplement, we provide information about the Small-Angle Neutron Scattering (SANS) data analysis (Section \ref{sec:sans}) and calculations of model form factors of spin vortices as seen by SANS (Section \ref{sec:vortices}). In Section \ref{sec:sans_t_and_h}, we give a brief review of the effects of temperature and cooling field on the observed SANS patterns. Mean-field modeling (Section \ref{sec:meanfield}), relevant to our weakly frustrated Ni$_{0.81}$Mn$_{0.19}$ sample, is also exposed. Finally, an extended comparative analysis of the internal structure of the vortex-like defects evidenced in this work and that of frustrated skyrmions is presented (Section \ref{sec:mc_sims}).

\section{Neutron scattering data analysis} 
\label{sec:sans}

Small-Angle Neutron Scattering (SANS) data has been treated following the usual strategy\cite{Brulet2007}, which we briefly summarize here. In a first step, we have to eliminate the extrinsic background arising from direct beam tail and sample environment. The is done by measuring the raw intensity $I_{\rm empty~holder}^{\rm raw}$ obtained with the empty sample holder placed at the sample position. The latter is then subtracted from the scattering pattern to be analyzed, yielding

\begin{equation}
	I_{\rm i}^{\rm corr} = I_{\rm i}^{\rm raw} - t_{\rm i} \cdot I_{\rm empty~holder}^{\rm raw} \quad ,
	\label{eq:bgsub}
\end{equation}

where $i$ denotes the studied sample (Ni$_{81}$Mn$_{0.19}$) or a calibrant (in the present case, a single crystal of pure Ni). Then, scattered intensities can be converted to cross sections \emph{on an absolute scale} using the expression

\begin{equation}
	\sigma_{\rm NiMn} (q) = \frac{I_{\rm NiMn}^{\rm corr} \cdot t_{\rm Ni} \cdot d_{\rm Ni} \cdot e_{\rm Ni}}{\langle I_{\rm Ni}^{\rm corr (q)} \rangle \cdot t_{\rm NiMn} \cdot d_{\rm NiMn} \cdot e_{\rm NiMn}} \cdot \sigma_{\rm Ni}^{\rm inc} \quad ,
	\label{eq:absscale}
\end{equation}

where $t$, $d$ and $e$ denote transmissions, atomic densities and thicknesses, respectively, while $\sigma_{\rm Ni}^{\rm inc} = \frac{5.2}{4\pi}~$barn$\cdot$sr$^{-1}$ is the incoherent scattering cross section of Ni (taken from Ref. \onlinecite{NeutronDataBooklet}).

As shown by the results presented in Fig. 4 of main text, this strategy allows obtaining cross sections which agree to within a few \% between the experiments performed on two different SANS spectrometers (D33 at Institut Laue Langevin\cite{Dewhurst2008} and PAXY at Laboratoire L\'eon Brillouin\cite{PAXYsheet}).

To determine the longitudinal and transverse cross sections, we use the configuration b) where the magnetic field H is in the detector plane and $\alpha$ is the angle $(\mathbf{q},\mathbf{H})$. We average the scattering map within two angular sectors 
of 60$^{\circ}$: sector 1 for $\alpha$ =0 ($\mathbf{q}$ // $\mathbf{H}$) and sector 2 for $\alpha =\pi/2$ ($\mathbf{q} \perp \mathbf{H}$) (see Figs. 2 and 3 of main text). Then we combine the intensities of the two sectors to deduce $\sigma_{\rm T}(q)$ and $\sigma_{\rm L}(q)$.  The intensities write:  
\begin{subequations}
\begin{align}
	\begin{split}
		\sigma_{\rm T}(q)+I_{\rm bg}(q)/2 &= \sigma(\mathbf{q},0) / 2 \quad 
		\label{eq:sigmaT}
	\end{split}\\
	\begin{split}
		\sigma_{\rm L}(q)+I_{\rm bg}(q)/2 &= \sigma(\mathbf{q},\pi/2) -  \sigma(\mathbf{q},0) / 2\quad		
	\end{split} 
\label{eq:sigmaLT}
\end{align}
\end{subequations}	

In a second step we discuss the origin of the intrinsic background term $I_{\rm bg}(q)$, which must be evaluated to isolate the vortex cross sections $\sigma_{\rm T}(q)$ and $\sigma_{\rm L}(q)$. This background term  originates from the sample itself and is quasi-isotropic in the detector plane. It is well accounted by the expression:

\begin{equation}
	I_{\rm bg}(q) = \frac{a}{q^{\rm p}} + b \quad .
	\label{eq:bg_def0}
\end{equation}

The first term $\frac{a}{q^{\rm p}}$ dominates in the low-$q$ region ($q \leq 0.025 \text{\AA}^{-1}$) and strongly increases down to $q \rightarrow 0$ following a power law.  This  small-angle scattering originates from long range nuclear and magnetic inhomogeneities, such as crystal dislocations or magnetic domain walls. The nuclear part could be estimated in principle by performing measurements at a very large fields when the whole magnetic contribution is negligible, but the magnetic part is always present and introduces a slight anisotropy in the background intensity. By fitting the low-q tail, we find a power-law exponent $p = 3.2$ in all cases. This value has been fixed in the course of data evaluation.  

\begin{figure*}[!ht]
		\centering
		\includegraphics[width=0.98\textwidth]{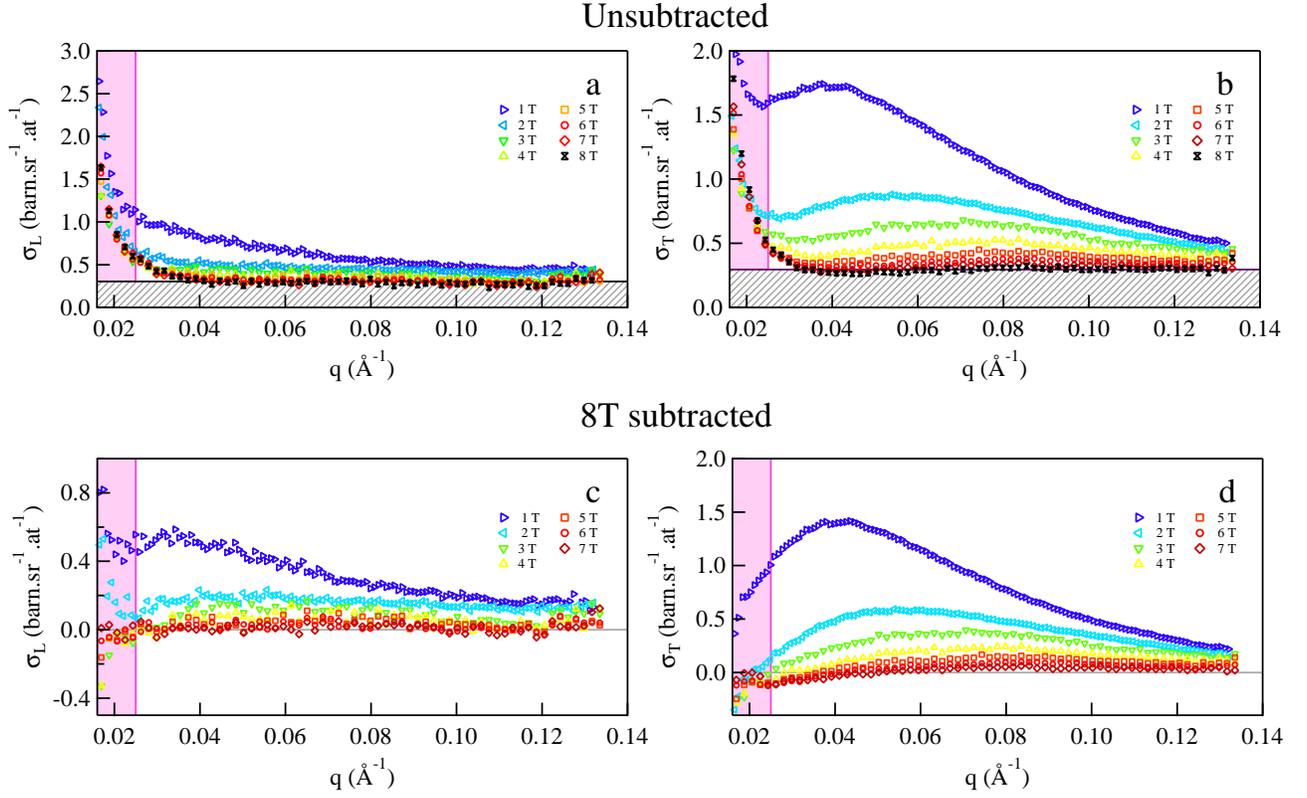}
		\caption{\label{fig:scalingSM}Evolution of the longitudinal \textbf{(a)} and transverse \textbf{(b)} magnetic neutron scattering cross sections as a function of magnetic field.  In panels  \textbf{(a)} and \textbf{(b)} (unsubtracted data), the hatched region shows the calculated constant background term $b$ and the pink shaded one is the region ($q \leq 0.025 \text{\AA}^{-1}$) where the low-$q$ tail $\frac{a}{q^{\rm p}}$ is important. In panels \textbf{(c)} and \textbf{(d)}, the spectrum measured at the maximum field of 8\,T is subtracted from each curve, to account for the intrinsic background term.}
\end{figure*}

The second term of Eq. \ref{eq:bg_def0} is a $q$-independent contribution arising from incoherent scattering and chemical disorder (namely the Laue scattering) in the sample.  It can be calculated exactly, knowing the amount of chemical short range order. This term is expressed as: 

\begin{widetext}
\begin{equation}
	b = \underbrace{(1-x) \cdot \sigma_{\rm Ni}^{\rm inc} + x \cdot \sigma_{\rm Mn}^{\rm inc}}_{\text{Incoherent scattering}} + \underbrace{x \cdot (1-x) \cdot \left( 1 + \sum_{\rm i = 1}^{3} z_{\rm i} \alpha(R_{\rm i}) \right) \cdot \left(b_{\rm Ni}^{\rm coh} - b_{\rm Mn}^{\rm coh}\right)^{2}}_{\text{Disorder scattering}} \quad ,
	\label{eq:bg_def1}
\end{equation}
\end{widetext}

where $x$ is the atomic concentration of Mn, $\sigma_{\rm Ni,Mn}^{\rm inc}$ the incoherent scattering cross sections of Ni and Mn, $b_{\rm Ni,Mn}^{\rm coh}$ the coherent scattering lengths of Ni and Mn, $z_{\rm i}$ the number of $i^{\rm th}$ neighbors in the face-centered cubic structure and $\alpha(R_{\rm i})$ the corresponding positional short-range order parameters obtained on Ni$_{\rm 0.8}$Mn$_{\rm 0.2}$\footnote{Although the results provided by these authors have been obtained on a sample with slightly different composition and heat treatment, we assume that their parameters are likely to be representative of the SRO of our Ni$_{\rm 0.81}$Mn$_{\rm 0.19}$ sample.} by Cable and Child\cite{Cable1974}  (see Tab. \ref{tab:sroparams}). Taken together, we obtain $b = 590$ mbarn. This yields a background level of $b/2 = 295$ mbarn for $\sigma_{\rm L}$ and $\sigma_{\rm T}$, in {\it quantitative} agreement with the experiment (see Fig. \ref{fig:scalingSM}, and Figs. 3e and 4a,b of main text). This excellent accordance shows that the corrections of environmental background and the calibration of the data in absolute scale are truly reliable.  


\begin{table}[!ht]
\caption{\label{tab:sroparams}Positional short-range order parameters of Ni$_{\rm 0.8}$Mn$_{\rm 0.2}$, from Ref. \onlinecite{Cable1974}.}
\begin{ruledtabular}
\begin{tabular}{lcc}
& $z$ & $\alpha(R)$\\
\hline
1$^{\rm st}$ neighbors & $12$ & $-0.09$ \\
2$^{\rm nd}$ neighbors & $6$ & $+0.07$ \\
3$^{\rm rd}$ neighbors & $24$ & $+0.02$ \\
\end{tabular}
\end{ruledtabular}
\end{table}

In summary, to account for the intrinsic background, we can either refine by fitting Eq. \ref{eq:bg_def0} to the data with the parameters $p$ and $b$ being fixed, or account for it by subtracting a pattern at the highest measured field of 8\,T. In Fig. \ref{fig:scalingSM} below, we compare the longitudinal and transverse cross sections with and without subtraction. The high field subtraction singles out the magnetic contribution from the vortices. However it is not very accurate at low $q$ ({\it i.e.} at $q \leq 0.025 \text{\AA}^{-1}$, see pink area in Fig. \ref{fig:scalingSM}), especially for the longitudinal cross section, and it neglects the vortex contribution at 8\,T, which can still be detected. This is why we have chosen the fitting procedure and we present the non subtracted data in the main text. 
We emphasize that the qualitative conclusions reached in the main text are unaffected by the choice of data treatment method. 

\section{Analytical expressions for the form factor of a spin vortex} 
\label{sec:vortices}

\begin{figure*}[!ht]
		\centering
		\includegraphics[width=0.98\textwidth]{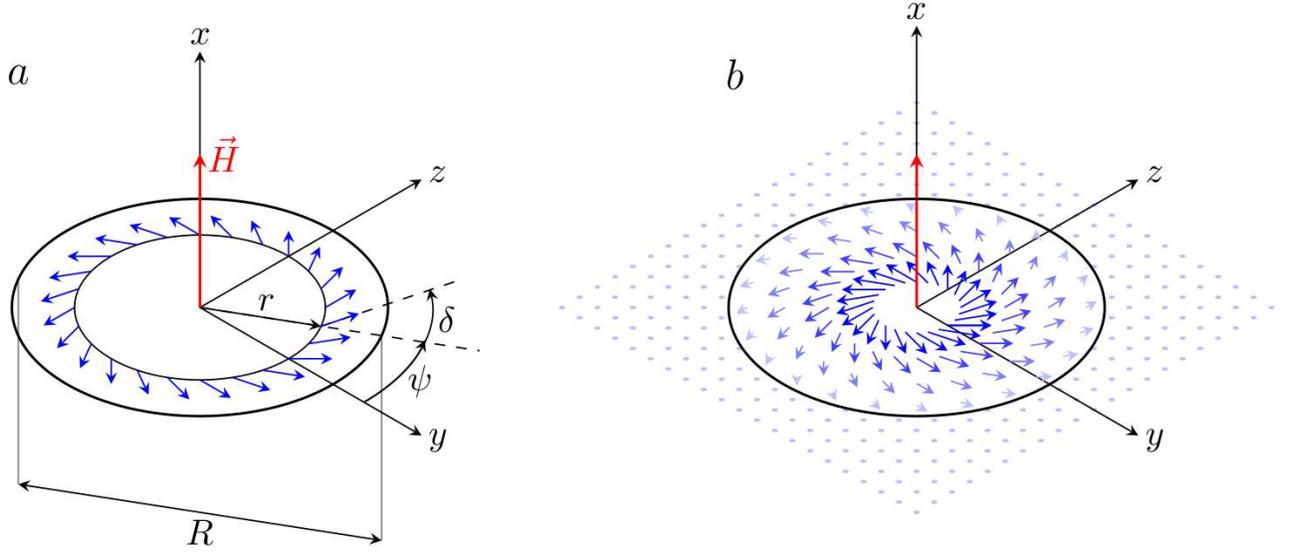}
		\caption{\label{fig:vortex} \textbf{(a)} Spin vortex corresponding to Eq. \ref{eq:m_rpsidelta} (see text). \textbf{(b)} Transverse 
		magnetization distribution, yielding the form factor given by Eq. \ref{eq:FFT_myz_smooth}.}
\end{figure*}

In order to support the results presented in main text, we carry out calculations of a spin vortex form factor as seen by SANS. We assume a regular vortex of radius R, in a field H//x  (see Fig.\ref{fig:vortex}). $m_{\rm L}(r$) and $m_{\rm T}(r)$ are the components of the local moment along the field and perpendicular to it, respectively called longitudinal and transverse.  Analytical expressions for the longitudinal $F_{\rm L}$ and transverse $F_{\rm T}$ form factors are obtained using \verb?Wolfram Mathematica 10.4?.\\ 

\vspace{11pt}
We start by defining the spin field that we use to model a regular vortex in the cartesian frame of Fig. \ref{fig:vortex}:

\begin{equation}
	\label{eq:m_rpsidelta}
	\mathbf{m} = \left( \begin{matrix}
m_{\rm L}(r) \\
m_{\rm T}(r) \cdot \cos \left(\psi+\delta\right) \\
m_{\rm T}(r) \cdot \sin \left(\psi+\delta\right)
\end{matrix} \right)
\quad ,
\end{equation}

where $r$ and $\psi$ are polar coordinates in the $(y,z)$-plane and $\delta$ the angle formed by individual spins with respect to the concentric vortex lines. The corresponding structure factors are obtained by Fourier transforming Eq. \ref{eq:m_rpsidelta}:

\begin{equation}
	F_{\rm i} = \frac{1}{\pi} \int_{-\pi}^{\pi} \left( \frac{1}{2 \pi R^2} \int_{-\pi}^{\pi} \int_{0}^{R} \, m_{\rm i} \, e^{i q r} \, r \, dr \, d\psi \right) \, d\delta \quad ,
	\label{eq:fft_m}
\end{equation}

where $i = \{x,y,z\}$. As expressed by Eq. 4 of the main text, the magnetic neutron scattering cross section explicitly contains $\langle F_{\rm i} \rangle^{2}$ and $\langle F_{\rm i}^{2} \rangle$. First neglecting a possible $r$-dependence of $m_{\rm L}$ and $m_{\rm T}$, symmetry considerations\footnote{In our choice of vortex morphology, $\mathbf{m}_{\rm L}$ is symmetric with respect to origin such that only the real part of its Fourier transform is non-zero. Conversely, the antisymmetric $\mathbf{m}_{\rm T}$-term simplifies to a sine Fourier transform.} lead to:

\begin{widetext}
\begin{equation}
	\label{eq:FFT_mx}
	\langle F_{\rm x}^{2} (q) \rangle = \langle F_{\rm x} (q) \rangle^{2} = \left(\frac{m_{\rm L}}{2 \pi R^2} \int_{0}^{R} \int_{-\pi}^{\pi} r \cos \left(q \, r \right) \, dr \, d\psi\right)^2 = \left(\frac{2 m_{\rm L}}{qR}\right)^{2} \cdot J_{\rm 1}^{2}(qR) \quad ,
\end{equation}

\begin{eqnarray}
\label{eq:FFT_my}
	\nonumber
	\langle F_{\rm y}^{2} (q) \rangle &=& \frac{1}{\pi} \cdot \int_{-\pi}^{\pi} \left(\frac{m_{\rm T}}{2 \pi R^2} \int_{0}^{R} \int_{-\pi}^{\pi} \cos \left(\psi + \delta\right)r \sin \left(q \, r \right) \, dr \, d\psi\right)^2 \, d\delta \quad\\
	\nonumber
	&=& \frac{1}{\pi} \cdot \int_{-\pi}^{\pi} \frac{\pi^2 \, m_{\rm T}^{2} \cdot \left(J_{\rm 1}(qR) \cdot H_{\rm 0}(qR) - J_{\rm 0}(qR) \cdot H_{\rm 1}(qR)\right)^{2} \cdot \cos^{2} \delta}{4 q^{2} R^{2}} \, d\delta\\
	&=& \frac{\pi^2 \, m_{\rm T}^{2} \cdot \left(J_{\rm 1}(qR) \cdot H_{\rm 0}(qR) - J_{\rm 0}(qR) \cdot H_{\rm 1}(qR)\right)^{2}}{4 q^{2} R^{2}} \quad ;\\
	\nonumber
	\quad\langle F_{\rm y} (q) \rangle^{2} &=& 0 \quad ,
\end{eqnarray}
\begin{eqnarray}
	\label{eq:FFT_mz}
	\nonumber
	\langle F_{\rm z}^{2} (q) \rangle &=& \frac{1}{\pi} \cdot \int_{-\pi}^{\pi} \left(\frac{m_{\rm T}}{2 \pi R^2} \int_{0}^{R} \int_{-\pi}^{\pi} \sin \left(\psi + \delta\right)r \sin \left(q \, r \right) \, dr \, d\psi\right)^2 \, d\delta \quad\\
	\nonumber
	&=& \frac{1}{\pi} \cdot \int_{-\pi}^{\pi} \frac{\pi^2 \, m_{\rm T}^{2} \cdot \left(J_{\rm 1}(qR) \cdot H_{\rm 0}(qR) - J_{\rm 0}(qR) \cdot H_{\rm 1}(qR)\right)^{2} \cdot \sin^{2} \delta}{4 q^{2} R^{2}} \, d\delta\\
	&=& \frac{\pi^2 \, m_{\rm T}^{2} \cdot \left(J_{\rm 1}(qR) \cdot H_{\rm 0}(qR) - J_{\rm 0}(qR) \cdot H_{\rm 1}(qR)\right)^{2}}{4 q^{2} R^{2}}\quad ;\\
	\nonumber
	\langle F_{\rm z} (q) \rangle^{2} &=& 0 \quad ,
\end{eqnarray}
\end{widetext}

where $J_{\rm n}$ ($H_{\rm n}$) are Bessel (Struve) functions of order $n$. We note that Eqs. \ref{eq:FFT_mx}, \ref{eq:FFT_my} and \ref{eq:FFT_mz} are equivalent to the results obtained by Metlov and Michels for a centered vortex in a ferromagnetic nanodot (see Eq. 13 from Ref. \onlinecite{Metlov2016}).\\

\begin{figure*}[!ht]
		\centering
		\includegraphics[width=0.98\textwidth]{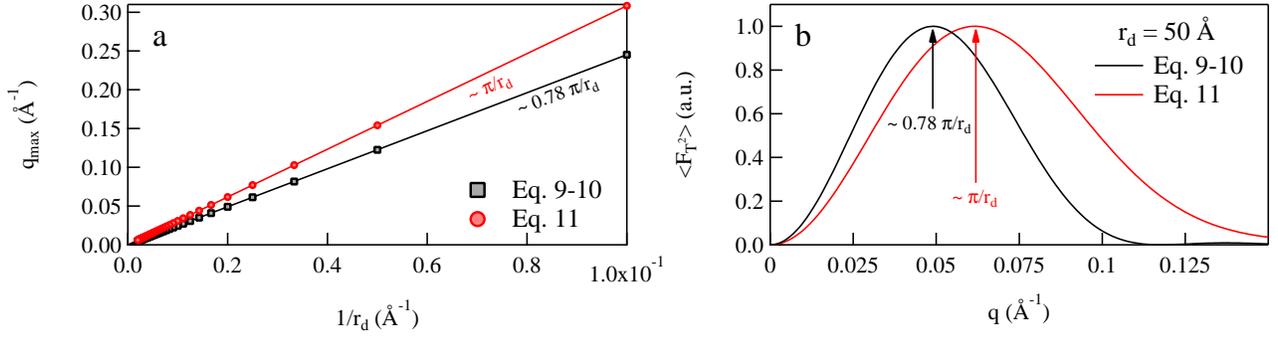}
		\caption{\label{fig:Fcalc} \textbf{(a)} Evolution of the magnetic scattering peak position as a function of the inverse defect radius for the two models described in text. In both case, the dependences are fully linear, yet with different slopes. \textbf{(b)} Transverse form factors for a vortex of 50 \AA~radius calculated using Eqs. \ref{eq:FFT_my}-\ref{eq:FFT_mz} and Eq. \ref{eq:FFT_myz_smooth}, respectively. In the second case, the peak shape is asymmetric, in agreement with the experimental SANS data.}
\end{figure*}

\vspace{11pt}
 The average squared form factor $\langle F_{\rm y}^{2} (q) \rangle =\langle F_{\rm z}^{2} (q) \rangle$ shows a maximum vs. q (Fig. \ref{fig:Fcalc}), as expected since the transverse spin components compensate within the vortex, yielding zero intensity at q=0. However, in the context of a bulk disordered ferromagnet, we cannot physically consider constant transverse magnetization since it would imply a sudden jump to $m_{\rm T} = 0$ outside of the vortex, {\it i.e.} in the average ferromagnetic medium. If we assume instead that $m_{\rm T} (r)$ is maximum at the vortex center ($r = 0$) and continuously decreases away from the center as $m_{\rm T}(r) = m_{\rm T} \cdot (1 - r/R)$ (see Fig. \ref{fig:vortex}), this choice restores continuity at the vortex edge. This average description neglects the local variations of the moments orientations  such as those induced by different Mn and Ni moments, or by an AF core constituted of first neighbor Mn-Mn pairs, which would yield smooth modulations of the diffuse scattering at larger q-values.  With these assumptions, the transverse form factors are now expressed as:

\begin{widetext}
\begin{eqnarray}
\label{eq:FFT_myz_smooth}
	\langle F_{\rm y}^{2} (q) \rangle &=& \langle F_{\rm z}^{2} (q) \rangle\\
	\nonumber
	&=& \frac{m_{\rm T}^{2} \cdot \left[J_{\rm 1}(q \, R) \cdot \left(\pi q \, R  \, H_{\rm 0}(q \, R) - 4\right) + q \, R \, J_{\rm 0}(q \, R) \cdot \left(2 - \pi H_{\rm 1}(q \, R)\right)\right]^{2}}{q^{4} R^{6}} \quad ;\\
	\nonumber
	\langle F_{\rm y} (q) \rangle^{2} &=& \langle F_{\rm z} (q) \rangle^{2} = 0 \quad .
\end{eqnarray}
\end{widetext}

As shown in Fig. \ref{fig:Fcalc}, Eqs. \ref{eq:FFT_my}-\ref{eq:FFT_mz} and Eq. \ref{eq:FFT_myz_smooth} both yield peaks in the transverse form factors, but at different q values. The peak positions $q_{\rm max}$ differ appreciably but have a $1/r_{\rm d}$-dependence in common: $q_{\rm max} \simeq 0.78 \, \pi / r_{\rm d}$ in the former case, $q_{\rm max} \simeq \pi / r_{\rm d}$ in the latter case (see Fig. \ref{fig:Fcalc}a). As explained above, the second model seems closer to reality for continuity reasons. This motivated our choice to relate the defect size to $\pi / q_{\rm max}$ (see main text). Eq. \ref{eq:FFT_myz_smooth} also yields more asymmetric peaks, in closer proximity to the experimental transverse cross sections (see Fig. \ref{fig:Fcalc}b and compare with Fig. 3e of main text). The variation of  $m_{\rm T}(r)$ should be accompanied with a correlated variation of $m_{\rm L}(r)$ with respect to the average magnetization, which is likely too small to be observed (see main text).  

\section{Effect of temperature and cooling-field on the SANS patterns}
\label{sec:sans_t_and_h}

While the work presented in main text is focused on the low temperature properties of the defects studied by small-angle neutron scattering and their relation to the ground state properties of Ni$_{0.81}$Mn$_{0.19}$, we provide here some details about the temperature- and cooling-field dependence of the experimental patterns. 

\begin{figure*}[!ht]
		\centering
		\includegraphics[width=0.98\textwidth]{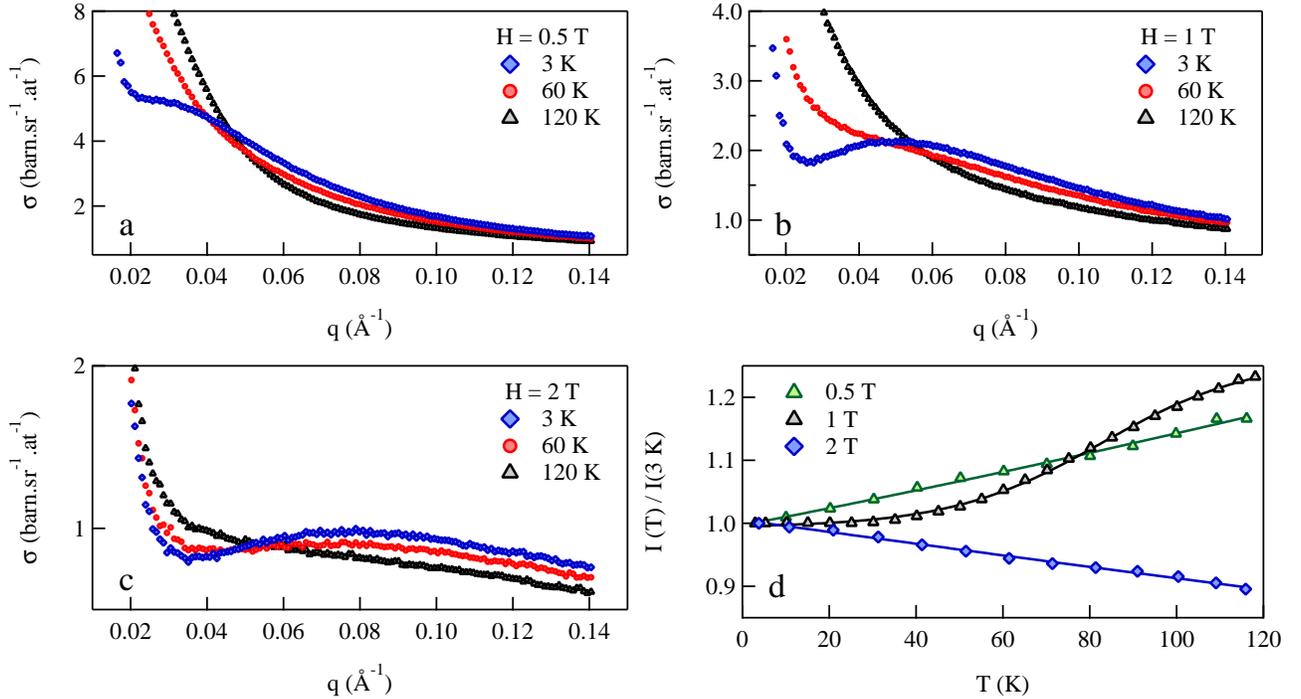}
		\caption{\label{fig:SANS_t_and_hcool1}\textbf{(a-c)} Temperature-dependence of the SANS patterns for applied fields $H = 0.5$ T \textbf{(a)}, $1$ T \textbf{(b)} and $2$ T \textbf{(c)} recorded upon cooling from T $\simeq$ T$_{\rm K} \sim 120$ K. \textbf{(d)} Temperature-dependence of the scattered intensity, integrated over the explored momentum transfer range $0.016 \leq q \leq 0.14 \text{\AA}^{-1}$ and scaled to its value at T = 3 K.}
\end{figure*}

\vspace{11pt}
As shown in Fig. \ref{fig:SANS_t_and_hcool1}a-c, increasing temperature leads to the progressive rise of the small-angle intensity and a concomitant vanishing of the peak feature in the observed cross section, already well-below the canting temperature T$_{\rm K} \sim 120$ K. This is due to the thermal activation of spin waves, which contribute to the scattered intensity since inelastic processes are not filtered out in a SANS setup. This leads to a non-trivial evolution of the total intensity, which strongly depends on the applied magnetic field (see Fig. \ref{fig:SANS_t_and_hcool1}d). It has been shown in Ref. \onlinecite{Lequien1987} (using a three-axis spectrometer which allows isolating purely elastic scattering at the expense of neutron flux) that the signal associated with vortex-like defects vanishes only at T$_{\rm K}$ while the vortex size remains basically constant as a function of temperature, in agreement with theoretical expectations. Altogether, this justifies our experimental strategy which concentrates on the low-temperature regime, where the properties of the field-induced "croutons" can be conveniently studied and compared to "T = 0" MC simulations.   

\begin{figure*}[!ht]
		\centering
		\includegraphics[width=0.98\textwidth]{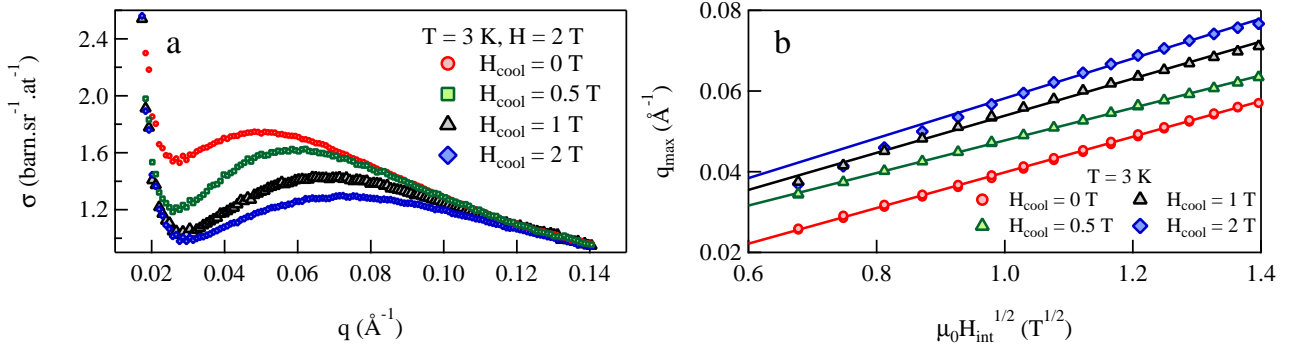}
		\caption{\label{fig:SANS_t_and_hcool2}\textbf{(a)} Influence of a cooling-field $H_{\rm cool}$ on the low temperature SANS patterns, showing a systematic shift and decrease of the peak position and intensity, respectively. \textbf{(b)} Field-dependence of the peak position $q_{\rm max}$ for different values of $H_{\rm cool}$ at T = 3 K.} 
\end{figure*}

\vspace{11pt}
On the other hand, in the Ni$_{0.81}$Mn$_{0.19}$ sample which behaves as a "rigid" system, the application of a magnetic field $H_{\rm cool}$ upon cooling reveals the existence of unidirectional anisotropy induced by Dzyaloshinskii-Moryia interactions, the cooling field acting as an additional bias field \cite{Ziq1990}. Our SANS study
reveals that the transverse correlations are substantially modified by $H_{\rm cool}$, although the bare plateau magnetization is basically not affected \cite{Mirebeau1988}. Indeed, SANS patterns recorded for different values of $H_{\rm cool}$ at low temperature clearly display different peak positions and intensities (see Fig. \ref{fig:SANS_t_and_hcool2}a). Essentially, increasing $H_{\rm cool}$ favors smaller defects at equal applied field value. The scaling laws which govern the evolution of the defect size are however preserved, as shown in Fig. \ref{fig:SANS_t_and_hcool2}b where the law $q_{\rm max} \propto \mu_{\rm 0}H_{\rm int}^{1/2}$  remains valid for all fields (except for applied fields smaller than 1 T for $H_{\rm cool} = 2$ T).  These aspects highlight the importance of the neutron probe to the study of fine magnetic features of RSGs. 



\section{Mean-field phase diagram and effective antiferromagnetic bond concentration}
\label{sec:meanfield}

The re-entrant spin-glass (RSG) phase has been studied theoretically by many authors. Here, we use the celebrated model of Gabay and Toulouse\cite{Gabay1981} to compare the mean-field phase diagram with the experimental one and estimate the effective antiferromagnetic (AFM) bond concentration of our Ni$_{0.81}$Mn$_{0.19}$ sample.\\

\begin{figure*}[!ht]
		\centering
		\includegraphics[width=0.98\textwidth]{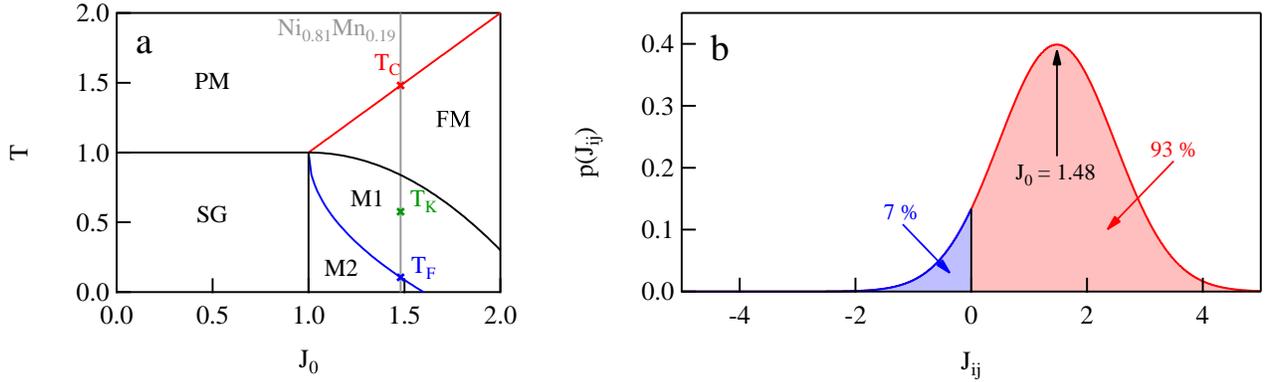}
		\caption{\label{fig:mf_gabaytoulouse} \textbf{(a)} Mean-field phase diagram of a bond-disordered ferromagnet calculated using the model of Gabay and Toulouse\cite{Gabay1981}. $T_{\rm C}$, $T_{\rm K}$ and $T_{\rm F}$ are the experimental Curie, canting and spin freezing temperature of Ni$_{0.81}$Mn$_{0.19}$, respectively. \textbf{(b)} Probability distribution function of the random-bond interactions calculated using Eq. \ref{eq:pdf_Jij} with $J_{\rm 0} = 1.48$. The blue (red) shaded region correspond to effective AFM (FM) interactions (see text).}
\end{figure*}

The phase diagram of RSG's (Fig. \ref{fig:mf_gabaytoulouse}) is calculated in mean field approximation for interactions with infinite range. The spins of the $\it{i}$ and $\it{j}$ sites interact via independent random interactions $J_{\rm ij}$, distributed according to  a normalized Gaussian law:

\begin{equation}
		p\left(J_{\rm ij}\right) = \sqrt{\frac{N}{2\pi}} \cdot \exp \left[-\frac{N}{2}\left(J_{\rm ij}-\frac{J_{\rm 0}}{N}\right)^{2}\right]
		\label{eq:pdf_Jij}
\end{equation}

where $N$ is the number of sites and $J_{\rm 0}$ is the average exchange interaction. Namely $\langle J_{\rm ij} \rangle_b = J_{\rm 0}/N$ and $\langle J_{\rm ij}^{2}  \rangle_b=1/N $  where $\langle \rangle_b $ denotes an average over bond disorder, that is over $p\left(J_{\rm ij}\right)$. 
 As shown in Fig. \ref{fig:mf_gabaytoulouse},
for $J_{\rm 0} = 0$ and $J_{\rm 0} \leq 1$, the low temperature phase is a spin glass (SG), showing no long range order. A tricritical point is observed for $J_{\rm 0} = 1$. In the region $J_{\rm 0} \geq 1$,  long range ferromagnetic order can occur. The system first evolves from paramagnetic (PM) to ferromagnetic (FM) at $T = T_{\rm C} = J_{\rm 0}$ upon cooling. At lower temperatures, two mixed phases, M1 and M2, are subsequently stabilized, corresponding to the freezing of transverse spin components (M1) and strong irreversibilities in the magnetization (M2). Most importantly, the magnetic LRO is not broken and persists both in M1 and M2 phases. This is the main difference between the RSG and SG's. In the usual terminology, the FM-M1 transition (or Gabay-Toulouse line) takes place at the canting temperature $T_{\rm K}$, while the M1-M2 transition (or de Almeida-Thouless line) occurs at the freezing temperature $T_{\rm F}$. We have kept these notations in the main text. The transition lines are calculated analytically according to Eqs. 10 and 11  of Ref. \onlinecite{Gabay1981}.

The Gabay-Toulouse model yields a rather accurate description of the experimental phase diagrams of RSG's systems by mapping the average $J_{\rm 0}$ interaction to the concentration of magnetic species. In Ni$_{0.81}$Mn$_{0.19}$,  the characteristic temperatures determined by susceptibility (this work, see Fig. 1 of main text) and neutron scattering (Ref. \onlinecite{Lequien1987}) are $T_{\rm C} = 257$\, K, $T_{\rm F} = 18$\, K and  $T_{\rm K}\simeq 120$\, K, yielding a ratio $T_{\rm F} / T_{\rm C}$ close to $0.07$.
 
 We first determine the appropriate value of $J_{\rm 0}$ for our compound. In Fig. \ref{fig:mf_gabaytoulouse}, the ratio $T_{\rm F} / T_{\rm C} = 0.07$ corresponds to $J_{\rm 0} = 1.48$. Using this value, we see that $T_{\rm K}$ lies below the calculated transition line. The agreement is however satisfactory, considering: i) the simplifying hypothesis made to calculate the mean field diagram; ii) that the determination of $T_{\rm K}$ is non trivial and the associated error bar ought to be large.

Next, we use the derived value of $J_{\rm 0}$ to calculate the corresponding probability distribution function (PDF) of random-bond interactions used for MC simulations, assuming a random distribution of AFM bonds (-J) of concentration c in a ferromagnetic medium (J). Integrating the PDF over negative $J_{\rm ij}$'s, we determine an equivalent AFM bond concentration of $\simeq 7$\%. This justifies the comparison between experiment and MC calculation for a weak concentration of AFM bonds (namely 5 \%).

\section{A comparative analysis of internal structure of frustrated skyrmions and vortex-like defects}
\label{sec:mc_sims}

In the present section, we give a comparative analysis of internal structure of so called "frustrated skyrmions" (see Ref. \onlinecite{Leonov2015} for further details) and vortex-like defects investigated in the present paper.
In both cases, the competing FM and AFM exchange interactions lead to the stability of particle-like states with non-trivial topology but showing different inherent properties.

To stabilize "frustrated skyrmions", we consider the following model with FM nearest-neighbour (NN) and AFM next-nearest-neighbour (NNN)  exchange interactions:
\begin{align}
E=
-J_1 \sum_{\langle i,j\rangle}\mathbf{m}_i\cdot\mathbf{m}_j+J_2 \sum_{\langle\langle i,j\rangle\rangle}\mathbf{m}_i\cdot\mathbf{m}_j
-h \sum_im_i^z.
\label{energy}
\end{align}
where $\langle i,j \rangle$ and $\langle\langle i,j\rangle\rangle$ denote pairs of NN and NNN spins of unit length, $\mathbf{m}_i$, respectively, and $J_1,J_2>0$. The third  term describes the interaction with the magnetic field parallel to  $z$ axis.
%
%
The stability mechanism is provided by the quartic differential terms (in general, by the terms with higher-order derivatives) that appear in the continuum version of Eq. (\ref{energy}) and allow to overcome the limitations imposed by the Derrick theorem \cite{Derrick1964}.
This mechanism is a reminiscence of the original mechanism of the dynamical stabilization proposed by Skyrme \cite{Skyrme1962}.

\begin{figure*}[tb]
\includegraphics[width=0.9\textwidth]{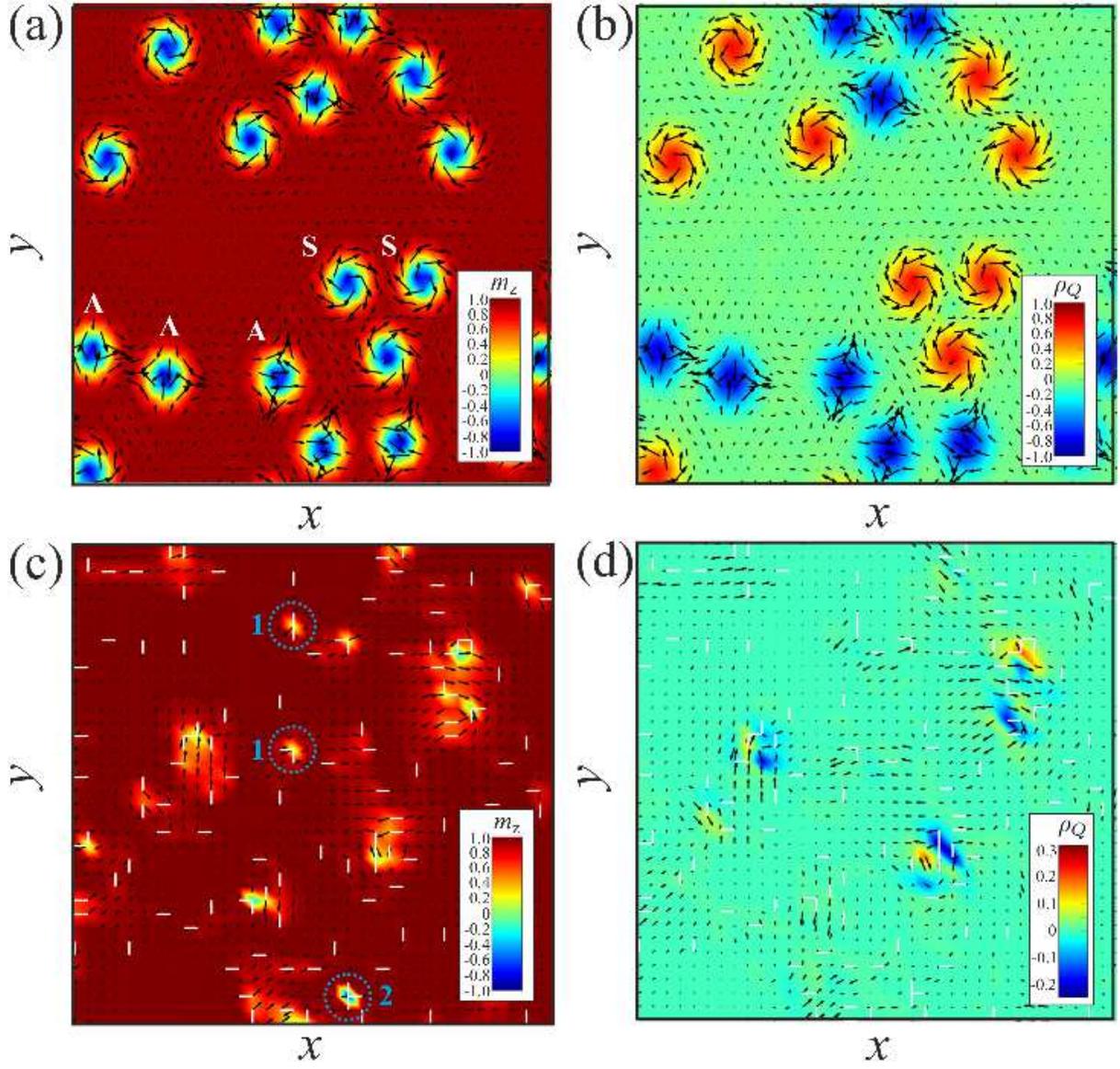}
\caption{
\label{skyrmionsSM} \textbf{(a)}, \textbf{(b)} Skyrmions (S) and antiskyrmions (A) in a frustrated ferromagnet with competing exchange interactions as described by the model (\ref{energy}) with $J_2/J_1=0.5,\,h=0.4$. Typical metastable states featuring clusters of skyrmions and antiskyrmions are obtained by relaxing the magnetic configuration with a random initial spin configuration. Color plots of $m_z$-components \textbf{(a)} and the topological charge density $\rho_Q$ \textbf{(b)} reflect the smooth rotation of the magnetization within the skyrmion cores. 
\textbf{(c)}, \textbf{(d)} Vortex-like defects induced by interaction disorder as described by the model (\ref{energy2}) with $c=0.05$ and $h=0.09$. Color plots of $m_z$-components \textbf{(c)} exhibit various types of spatially localized objects with the balanced topological charge density \textbf{(d)}. Blue numbered circles show vortices residing on two or three AFM bonds with distinct magnetization distributions (see text for details).
}
\end{figure*}

In "frustrated skyrmions", the vector $\mathbf{m}$ is antiparallel to $z$-axis at the center and gradually rotates towards the field-aligned state  at the boundary, thus, resulting in circular particle-like states (Fig. \ref{skyrmionsSM}a,b).
The angle between two adjacent spins within the skyrmion cores is controlled by the ratio $J_2/J_1$ in Eq. (\ref{energy}) and in the present case ($J_2 / J_1 = 0.5$) may reach the value $\pi/3$.

A prominent property of such frustrated magnets with competing exchange interactions is that a skyrmion and an antiskyrmion have the same energy irrespective to their helicity.
Skyrmions and antiskyrmions are distinguished conventionally based on the sign of their topological charge being either $+1$ or $-1$, correspondingly. 
The topological charge density, $\rho_Q$, maintains the same sign within the skyrmion cores (see Fig. \ref{skyrmionsSM}b with orange and blue coloring used for skyrmions and antiskyrmions, respectively).

With an increasing magnetic field, the skyrmions may undergo the collapse into the homogeneous state. Within the discrete model (\ref{energy}), the collapse occurs when the negative $m_z$-component of the magnetization within the skyrmion cores reaches 0; at this point, a skyrmion 
can be abruptly unwound (see in particular Fig. 6c in Ref. \onlinecite{Leonov2016} for more details of such a process).

In some range of applied magnetic field, the skyrmions represent metastable states: their energy is higher than the energy of the homogeneously magnetized state.
Usually, as shown in Fig. \ref{skyrmionsSM}a, one observes clusters of such metastable skyrmions and anti-skyrmions with mutual attracting interaction that are embedded into the homogeneously magnetized matrix \cite{Leonov2015}.
With decreasing magnetic field, the skyrmions tend to crystallize predominantly in a hexagonal lattice with the densest packing of skyrmions: at some critical field ($h < 0.35$ in the model (\ref{energy}) with the chosen parameters \cite{Leonov2015}), the energy of an isolated skyrmion becomes negative with respect to the homogeneous state and hence the skyrmions try to fill the space. 
The extended skyrmion lattice is determined by the stability of the localized solitonic skyrmion cores and their geometrical incompatibility in the corners of hexagons which frustrates regular space-filling.

However if the formation of a skyrmion lattice is suppressed, isolated skyrmions continue to exist below the critical field.
At the same time, isolated skyrmions have a tendency to elongate and expand into a band with helicoidal or cycloidal modulations and eventually to fill the whole space, since the spiral state represents the minimum with lower energy as compared to the local minima with the metastable isolated skyrmions.

Thus, the existence region of "frustrated" skyrmions is restricted by a collapse at high fields and the critical low field at which the energy of an isolated skyrmion becomes negative and thus instigates the formation of the lattice or its elliptical instability \cite{Leonov2016}.

The processes of lattice formation or elliptical instability are obviously not the case for vortex-like defects: residing around the AFM bonds, the vortices cannot form any type of an extended ordered phase. 
However, as pointed out in the main text, with the decreasing magnetic field and an increasing concentration of AFM bonds, the vortices form a liquid -like order stipulated by the same tendency to fill the space although remaining metastable solutions.

Solutions for vortex-like defects induced by interaction disorder are  obtained by minimizing the following Hamiltonian:

\begin{align}
E=
- \sum_{\langle i,j\rangle}J_{ij}\mathbf{m}_i\cdot\mathbf{m}_j
-h \sum_im_i^z.
\label{energy2}
\end{align}
where the sum $\langle i,j \rangle$ runs only over NN pairs. The $J_{ij}$ are independent random variables taking the values $+1$ and $-1$ with probability $1-c$ and $c$, correspondingly (see Sec. \ref{sec:meanfield}). 
The method used to obtain the vortex configurations in Fig. \ref{skyrmionsSM}c,d is described in Ref. \onlinecite{Kawamura1991} and is basically the same as for "frustrated" skyrmions\cite{Leonov2015}.

The stabilized vortices are metastable solutions in the whole range of an applied magnetic field. 
The vortices do not bear any smooth rotation of the magnetization and do not have any preferred helicity. 
As a consequence, the topological charge density has both signs within one isolated vortex as depicted in Fig. \ref{skyrmionsSM}d. 
The largest angle value between two adjacent spins obviously may reach $\pi$ (see, in particular, a vortex encircled by a blue dotted line and numbered 2 in Fig. \ref{skyrmionsSM}c). 
The common case, however, are the spatially localized vortices with the positive $m_z$-component (which is impossible for frustrated skyrmions) and rather small angles between spins (see vortices numbered 1 in Fig. \ref{skyrmionsSM}c). 
Having different number of AFM bonds within the cores, the vortices also exhibit different collapse fields. 
In particular, for the configuration depicted in Fig. \ref{skyrmionsSM}c, no vortices rest on single AFM bonds, and vortices on two bonds (blue circles) are close to their transformation into the homogeneous state.  
Thus, a specified internal structure of a vortex might reflect an exact distribution of AFM bonds in its core or its vicinity, which allows classification of such vortex states with the subsequent engineering of their properties. 
In particular, an endevour with creating a smooth magnetization rotation by a particular pattern of AFM bonds could  be undertaken. 
Thus, the existence region of vortex-like defects is restricted by a collapse at high fields (that depends on the internal structure of isolated particles) and the critical low field (and/or concentration $c$) at which the vortices form a liquid-like state. 

\begin{figure*}[!ht]
		\centering
		\includegraphics[width=0.98\textwidth]{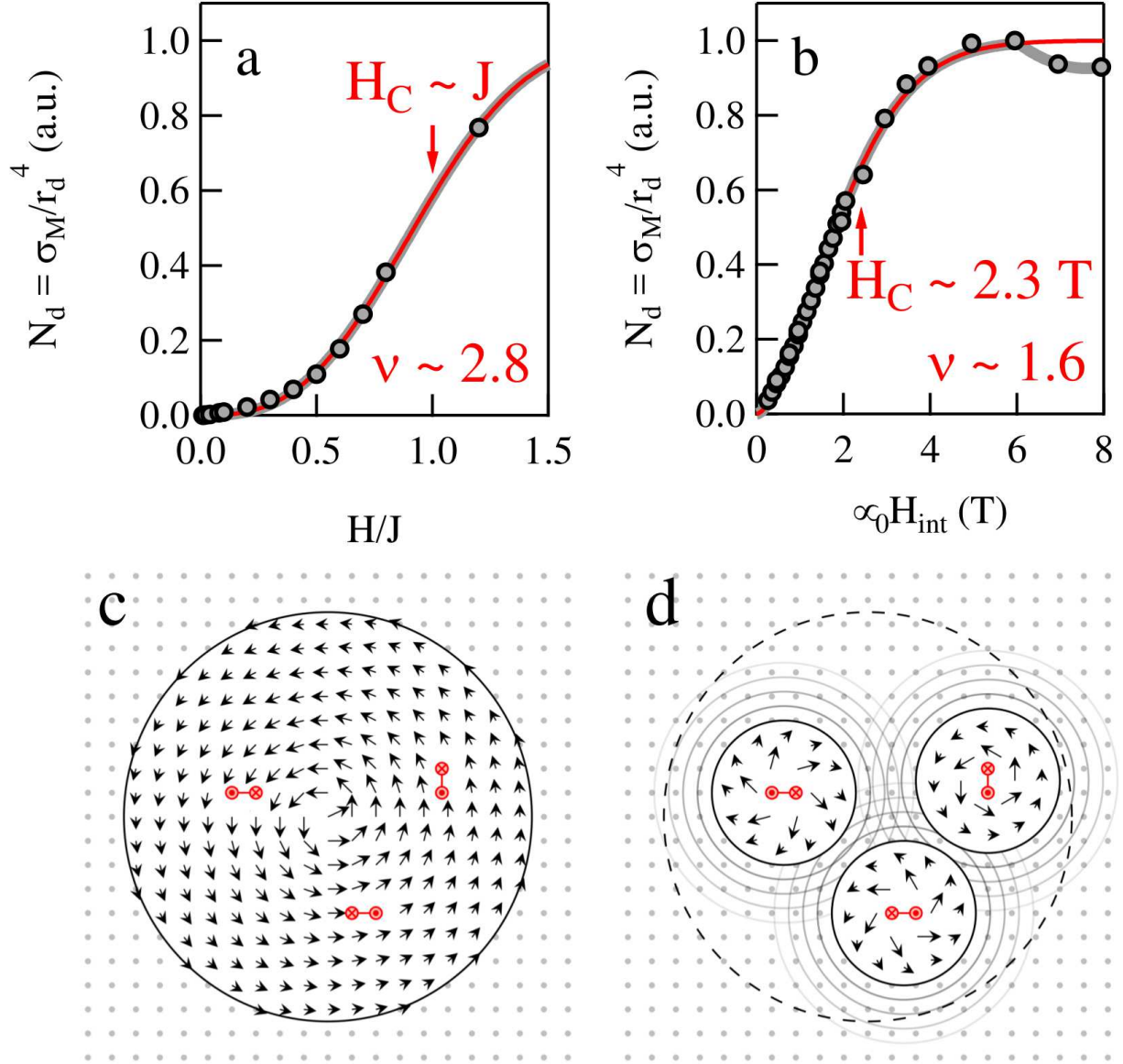}
		\caption{\label{fig:nd_vs_h_simvsexp}  \textbf{(a,b)} Field-dependence of the number of isolated vortex-like defects $N_{\rm d}$ as seen by MC simulation \textbf{(a)} and small-angle neutron scattering \textbf{(b)}. In both panels, red lines represent fits of Eq. 4 of main text to the data. \textbf{(c,d)} Schematic evolution of the vortex-like defects. \textbf{(c)} At low magnetic fields, large vortices encompass several AFM bonds (shown in red). \textbf{(d)} With increasing field, they shrink and become singled out, leading to the observed increase of $N_{\rm d}$ in \textbf{(a,b)}.}
\end{figure*}

Finally, it is instructive to compare the field-dependence of the number of isolated vortex-like defects $N_{\rm d}$ as obtained by MC simulation with that derived from our small-angle neutron scattering experiment (see Figs. 4f and 6f of main text, and Fig. \ref{fig:nd_vs_h_simvsexp}). 
As explained in main text, a qualitative agreement between calculation and experiment is observed, namely a global increase of $N_{\rm d}$ with increasing field, followed by a saturation at a finite field. 
Remarkably, a fit of a stretched exponential (Eq. 4 of main text) to the data yields similar agreement in the two cases (Fig. \ref{fig:nd_vs_h_simvsexp}a,b). 
This underscores similar evolutions of the defect size and integrated scattering intensity (or square amplitude of the Fourier transform of the transverse magnetization distribution), despite seemingly different local interaction schemes (see main text). However, we point out that the inflection point of the $N_{\rm d}$ \emph{vs.} $H$ occurs around $H_{\rm C} \simeq J$ in the simulation. A mapping to the experimental case suggests that $N_{\rm d}$ should not cease increasing for fields much smaller than several 100 T in our Ni$_{0.81}$Mn$_{0.19}$ sample, as opposed to our observations. The origin of such disagreement might be due to the willingly simplified approach we have followed to model the distribution of magnetic interactions in the MC simulations. In order to achieve a better control on the properties of systems supporting vortex- or skyrmion-like textures, our work could thus motivate further theoretical work, combining the effects of exchange frustration\cite{Okubo2012,Leonov2015,Lin2016} and quenched disorder\cite{Chudnovsky2017}, which have only been considered separately up to now. However, this will not be a simple exercise.

Altogether, our combined experimental and numerical investigation of the properties of the field-induced vortex-like defects stabilized in the Ni$_{0.81}$Mn$_{0.19}$ RSG leads to a simple physical picture. At low magnetic fields (Fig. \ref{fig:nd_vs_h_simvsexp}c), the defects are large and encompass several AFM (Mn-Mn) pairs, such that their apparent number $N_{\rm d}$ is small. Upon field increase, however, their size decrease while they remain fixed around the AFM pairs (Fig. \ref{fig:nd_vs_h_simvsexp}d), the number of which being fixed by the Mn concentration and heat treatment. This leads to an increase of $N_{\rm d}$. Its value is expected to saturate when defects start collapsing, {\it i.e.} when spins are locally aligned by the applied field.  


\end{document}